\documentclass[aps,prl,twocolumn,amsmath,amssymb,showpacs,longbibliography]{revtex4-1}

\usepackage{ifpdf} 
\usepackage[sort&compress]{natbib}
\usepackage{color}
\usepackage{soul}
\usepackage{commath}
\usepackage{amsmath,latexsym}

\ifx\pdfoutput\undefined 

\usepackage{graphicx} 
\else 
\usepackage[pdftex]{graphicx} 
\usepackage{epstopdf} 
\fi 
\usepackage[normalem]{ulem}

\preprint{}

\begin{document}

\title{Fragile-to-Strong Crossover, growing length scales, and dynamic heterogeneity in Wigner Glasses}

\author{Hyun Woo Cho$^{1}$, Mauro L. Mugnai$^{1}$, T. R. Kirkpatrick$^{2}$ and D. Thirumalai$^{1}$} 
\affiliation{$^{1}$Department of Chemistry, University of Texas at Austin, Austin, Texas 78712, USA}
\affiliation{$^{2}$Institute for Physical Science and Technology, and Department of Physics, University of Maryland, College Park, Maryland 20742, USA}

\date{\today}

\begin{abstract}

Colloidal particles, which are ubiquitous, have become ideal testing grounds for the structural glass transition (SGT) theories. In these systems glassy behavior is manifested as the density of the particles is increased.   Thus, soft colloidal particles with varying degree of softness capture diverse glass forming properties, observed normally in molecular glasses.  By performing Brownian dynamics simulations for a binary mixture of micron-sized charged colloidal suspensions, known to form Wigner glasses, we show that by tuning the softness of the interaction potential, achievable by changing the monovalent salt concentration, there is a continuous transition between fragile to strong behavior.  Remarkably, this is found in a system where the well characterized interaction potential between the colloidal particles is isotropic. We also show that the predictions of the random first order transition (RFOT) theory quantitatively describes the universal features such as the growing correlation length, $\xi\sim\Big(\frac{\phi_K}{\phi}-1\Big)^{-\nu}$ with $\nu = \frac{2}{3}$ where $\phi_K$, the analogue of the Kauzmann temperature, depends on the salt concentration. As anticipated by the RFOT predictions, we establish a causal relationship between the growing correlation length and a steep increase in the relaxation time and dynamic heterogeneity as the system is compressed. The broad range of fragility observed in Wigner glasses, which can be induced by merely changing the salt concentration, is used to draw analogies with molecular and polymer glasses.  The large variations in the fragility is found only when the temperature dependence of the viscosity is examined for a large class of diverse glass forming materials. In sharp contrast, this is vividly illustrated in a single system that can be experimentally probed. Our work also shows that the RFOT predictions are accurate in describing the dynamics over the entire density range, regardless of the fragility of the glasses, implying that the physics describing the structural glass transition is universal. 

\end{abstract}

\flushbottom
\maketitle

\section{introduction}

The abiding interest in the structural glass transition (SGT) problem, which occurs readily in a large class of materials either by supercooling or by compression, is a testimony to its importance in condensed matter physics. Extensive experimental, theoretical, and computer simulation studies have established that the Random First Order Transition (RFOT) theory~\cite{Kirkpatrick89PRA} provides a reasonable description of many important characteristics of the SGT. Several reviews~\cite{Kirkpatrick95,Lubchenko07ARPC,Parisi10RMP,Berthier11RMP, Biroli12Book,Kirkpatrick15RMP} have discussed the theoretical underpinnings and applications of the RFOT theory to not only to SGT but a number of other fields~\cite{Kirkpatrick15RMP}. Although the RFOT theory was inspired by a class of precisely soluble mean field spin glass models lacking inversion symmetry in the presence of quenched randomness~\cite{Kirkpatrick87PRL,Kirkpatrick87PRB,Kirkpatrick87PRBa,PhysRevB.37.5342,PhysRevB.38.4881}, it was shown that the major results could also be derived using a density functional Hamiltonian for liquids where the randomness is self-generated~\cite{Kirkpatrick89JPhysA}, just as in the SGT (see also \cite{Bouchaud94JP,Franz95PRL}). Because the theoretical approaches were inherently mean field-like, which although one could argue is accurate in liquids, additional studies were needed to assess the robustness of the RFOT conclusions. There are connections between equilibrium and dynamical transitions in large dimensions~\cite{Kirkpatrick97PRA}, which are explicit in spin glass models without inversion symmetry.   These have made precise in a number of remarkable studies~\cite{Kurchan12JST,Kurchan13JPCB,Charbonneau14JST}, which established that RFOT is exact in $d= \infty$ dimensions for hard sphere glass forming systems.  These studies provide support to the original suggestion~\cite{Kirkpatrick89PRA,Kirkpatrick89JPhysA} that the physics underlying RFOT describes the SGT problem fairly accurately. 

A prediction of the RFOT is that for a generic glass forming system there are two major transitions as the liquid is compressed (increase in the volume fraction of the particles, $\phi$). We focus on $\phi$, and not the temperature because that is the the relevant variable in the binary mixture of charged colloidal suspensions, which undergo a liquid to glass (Wigner glass) transition at high enough values of $\phi$~\cite{Lindsay:1982gn}.  At a $\phi \sim \phi_d$ there is a dynamical transition at which the transport starts to become sluggish although signatures of slow dynamics is evident even at values of $\phi$ less than $\phi_d$. As the liquid is compressed further ($\phi$ is increased) there is an ideal glass transition at $\phi_K$ (analogue of the Kauzmann temperature) at which the configurational entropy vanishes, which in turn results in complete cessation of motion. If undercooled by lowering the temperature instead of increasing $\phi$, the ideal glass transition occurs at $T_K$ where there is an essential singularity in the temperature dependence of the viscosity. Of course, the transition at $\phi_d$ and the thermodynamic transition at $\phi_K$ are connected, which is needed to provide a consistent picture of the SGT~\cite{Kirkpatrick95,Kirkpatrick15RMP}. The  topology of the state space is unremarkable at $\phi < \phi_d$ where collective transport is not prominent. On the other hand, for
$\phi > \phi_d$ the dynamics slows down because the system is trapped in one of the exponentially large number of 
metastable states~\cite{Kirkpatrick89JPhysA}. Under these circumstances transport becomes possible only by overcoming free energy barriers separating the metastable states. The time scales for crossing the barriers  can be arbitrarily long depending on $\phi$, and becomes essentially infinite at $\phi_K$. The two transition picture and the associated scaling relations of quantities, such as the growing length scales and the surface tension between two mosaic states at values of $\phi > \phi_d$ (or $T<T_d$), approaching the ideal glass transition volume fraction $\phi_K$ (or $T_K$), have been measured both in computer simulations using predominantly hard spheres (HS) or Lennard-Jones (LJ) or soft sphere (SS) mixtures~\cite{Thirumalai93PRE,Mountain87PRA,Kob:2012in,Ozawa6914,Berthier:2017du,Biroli:2008jy,PhysRevLett.114.205701} and experiments~\cite{Nagamanasa15NatPhys,Ganapathi18NatComm,Gokhale16AdvPhys,Albert:2016fm}.  

One of the hallmarks of glass forming materials  is that they exhibit dramatically different curvatures when $\log\eta$ is plotted as a function of $\frac{T_g}{T}$ where $\eta$ is the shear viscosity, $T_g$ is the glass transition temperature, which is operationally defined using $\eta(T_g) \simeq10^{13}$ poise or when the structural relaxation time reaches about 100 seconds. In the graph of $\log \eta$ as a function of $\frac{T_g}{T}$, often referred to as the Angell plot~\cite{Angell95Science}, classic glass formers, such as {\it ortho}-terphenyl or Trehalose are ``fragile" as are mixtures of HS, LJ or SS particles. In contrast, Si or SiO$_2$, which are network forming materials with anisotropic interaction potentials, are classified as ``strong" glasses. In fragile glasses, the effective activation free energies separating the metastable states explored above $\phi_d$ increase sharply as the system is continuously compressed whereas they are relatively independent of $\phi$ in strong glasses. Fragile and strong glasses are often discussed in terms of the fragility index~\cite{Bohmer:1993ce}. 

Here, we have two goals in mind. First, we demonstrate using mixtures of glass forming highly charged micron-sized colloidal suspensions (classical Wigner glasses) that the key predictions of RFOT are quantitatively validated, adding to the growing evidence that RFOT theory seamlessly explains the dynamics {\bf both} below $\phi_d$ and also in the density  range spanning $\phi_d \le \phi \le \phi_K$.  Second, we show that there is a crossover from fragile to strong behavior in Wigner glasses as the concentration of monovalent salt is increased. The large change in the fragility index needed for the crossover occurs in just one system even though the interparticle potential is isotropic. These new predictions can be tested using optical microscopy techniques~\cite{Gokhale16AdvPhys}.

The key to our findings is the recognition that stiffness of the interparticle potentials in colloids can be changed by controlling the surface charge or internal elasticity of colloids~\cite{Royall:2013bk}. Examples of such systems include emulsions~\cite{Nagamanasa15NatPhys}, microgels~\cite{Mattsson:2009jv,vanderScheer:2017hd,Philippe:2018ee}, charged colloids~\cite{Philippe:2018ee}, and squishable cells~\cite{PhysRevX.8.021025,Angelini4714}. These systems display glass-like properties that are distinct compared to fragile hard-sphere like systems. Most striking impact of the softness on the glass transition is that ``fragility" of colloids can be greatly modified upon change in the stiffness of the potential, which in turn can be altered by changing the interaction potentials~\cite{Mattsson:2009jv}. For the much less investigated Wigner glasses, the fragility is a measure of how steeply the relaxation time $\tau_\alpha$ increases near the glass transition volume fraction $\phi_g$. Since the first experiments using microgels as soft glasses~\cite{Mattsson:2009jv}, several studies revealed that the glass transition of soft colloids can be either strong or fragile depending on the stiffness~\cite{SeekellIII:2015cd,Nigro:2018wb,C7SM00739F,Yang_2010}. Upon decreasing the stiffness of the interaction potential between the colloids the fragility increases. Thus, soft colloids can be exploited as important model systems that mimic the characteristics of diverse glass forming materials~\cite{Angell:2009gx}. 

Despite several examples, the physics underlying the fragile-to-strong crossover in soft colloids is not fully understood. First, whether the softness does really contribute to the drastic change in the fragility remains controversial. Indeed, in several experiments~\cite{Philippe:2018ee,SahaD2015,vanderScheer:2017hd,C6SM02408D}, it was found that the fragility in soft colloids is insensitive to the softness of the interaction potentials. Previous simulations with model soft colloids showed that the drastic variation in $m_k$ cannot be reproduced by merely modifying the softness of the potential~\cite{Ninarello:2017bq,Shi:2011hc,Michele:2004bs,Philippe:2018ee}. Thus, some have argued that the softness does not dominate the fragility of colloidal glasses. Instead, it was suggested that other mechanisms relying on the the microscopic details of the soft colloids are important for the drastic change in the fragility in the previous experiments~\cite{Philippe:2018ee,Asai:2018jc,Gnan:2019ek}. Here, we elucidate the effect of the softness on the fragility of soft colloids in order to resolve the conflicting interpretations.        

Second, as a simplest realization of the diverse glass forming liquids, whether the glass transition in soft colloids with a broad spectrum of the fragility can be described universally is a question of fundamental importance. The sluggish dynamics near the glass transition is attributed to the sudden increase in the effective free energy barriers controlling structural relaxation at $\phi > \phi_d$. Thus, the fragility of liquids depends on how steeply the effective free energy barrier increases near the glass transition. As alluded to above, the RFOT theory naturally explains  the increase in the free energy barrier near the glass transition, which is due  to the emergence of a growing length scale in which dynamics of the particles are highly correlated \cite{Kirkpatrick89PRA,Kirkpatrick15RMP,Kirkpatrick:2014ud}. In the RFOT  the effective free energy barrier is characterized by a diverging length scale associated with the amorphous order \cite{Bouchaud:2004kk,Biroli:2008jy,Berthier:2012jg,Ganapathi18NatComm} or the correlated dynamics \cite{Flenner:2015go,Flenner:2011ht,Flenner:2014ck,Flenner:2010io}. According to the RFOT, therefore, regardless of the fragility of liquids, a significant increase in $\tau_\alpha$ should be universally described in terms of the growth of length scales. Testing this prediction of RFOT using Wigner glasses as an example of soft glasses with a broad range of tunable fragility is also an important motivation of this work. 

\section{Methods}

\textbf{Interaction potential:} Nearly four decades ago, Lindsay and Chaikin showed that increasing the volume fraction of binary mixtures of highly charged micrometer-sized colloidal particles results in Wigner glass formation, characterized by the absence of long-range order but with finite shear modulus \cite{Lindsay:1982gn}. Following our previous studies~\cite{Rosenberg_1989,Kang:2013kp}, we model the experimentally probed system as a mixture of charged spheres. The total number of the particles is $N=N_1+N_2$, where $N_1$ and $N_2$ are, respectively,  the number of small and large colloids. The bare radii of the particles, $a_1$ and $a_2$, are taken to be 0.525 $\mu\text{m}$ and 1.1 $\mu\text{m}$, corresponding to the ones used in the experiments. In our simulations, we choose $N_1=N_2=5000$. The interaction between the charged colloids is modeled by the Derjaguin-Landau-Verwey-Overbeek (DLVO) potential~\cite{Alexander:1984bj,Rosenberg:1987wk,Sanyal:1995kv,Thirumalai:1989il,Fisher:1994hi}, which is given by,
\begin{equation} \label{DLVO}
V_{ij}(r)=\frac{e^2Z_iZ_j}{4\pi\epsilon}\Bigg(\frac{\exp[\kappa a_i]}{1+\kappa a_i}\Bigg)\Bigg(\frac{\exp[\kappa a_j]}{1+\kappa a_j}\Bigg)\frac{\exp[-\kappa r]}{r}.
\end{equation}  
In Eq~(\ref{DLVO}), $Z_i$ is the valence of the charged colloids, whose values are 300 and 600 for small ($i=1$) and large ($i=2$) colloids, respectively, $r$ is the inter particle distance, and $\epsilon$ is the dielectric constant ($\epsilon=\epsilon_0\epsilon_r$, where $\epsilon_0$ and $\epsilon_r$ are vacuum and relative permittivity, respectively). Because  the charged colloids are solubilized in water at temperature $T=298K$, we use $\epsilon_r=78$. The influence of the counterions and the added monovalent salt on the interaction between the charged colloids is implicitly reflected in the inverse Debye-H\"{u}ckel screening length $\kappa$, given by,
\begin{equation} \label{kappa}
\kappa^2=\frac{e^2}{\epsilon k_B T}\Bigg(\rho_c z_c^2+\sum_{i'}^n \rho_{i'} z_{i'}^2\Bigg),
\end{equation} 
where $\rho_c$ and $z_c$ are the number density and valence of the counterions, and $k_B$ is the Boltzmann constant. For monovalent ions $|z_c|=1$, and therefore due to charge neutrality $\rho_c$ is given by $\rho_c=\rho_1Z_1+\rho_2Z_2$, where $\rho_1$ and $\rho_2$ are the number densities of the small and large colloids, respectively. In Eq~(\ref{kappa}), $\rho_{i'}$ and $z_{i'}$ are the number density and valence of the added salt, respectively. For monovalent ions, $\sum_{i'}^n \rho_{i'} z_{i'}^2$ becomes $\rho_{add}=\sum_{i'}^{n=2} \rho_{i'}$. We define the relative number density of the excess ions as, $\rho_r=\rho_{add}/\rho_c$, and consider values of $\rho_r$ ranging from 0 to 10. For simulation efficiency, $V_{ij}(r)$ is truncated and shifted at $r_{cut}$ where $V_{ij}(r_{cut})=0.001k_BT$. Because the interactions in Eq~(\ref{DLVO}) are screened, it is not necessary to use Ewald sums, which is usually required for simulating systems with particles interacting by long-ranged Coloumb interactions. We investigated the dynamics  of the charged colloids was investigated by carrying out extensive Brownian dynamics (BD) simulations, which we describe below (see Appendix A).  

\begin {figure}
\centering
\includegraphics  [width=3.5in] {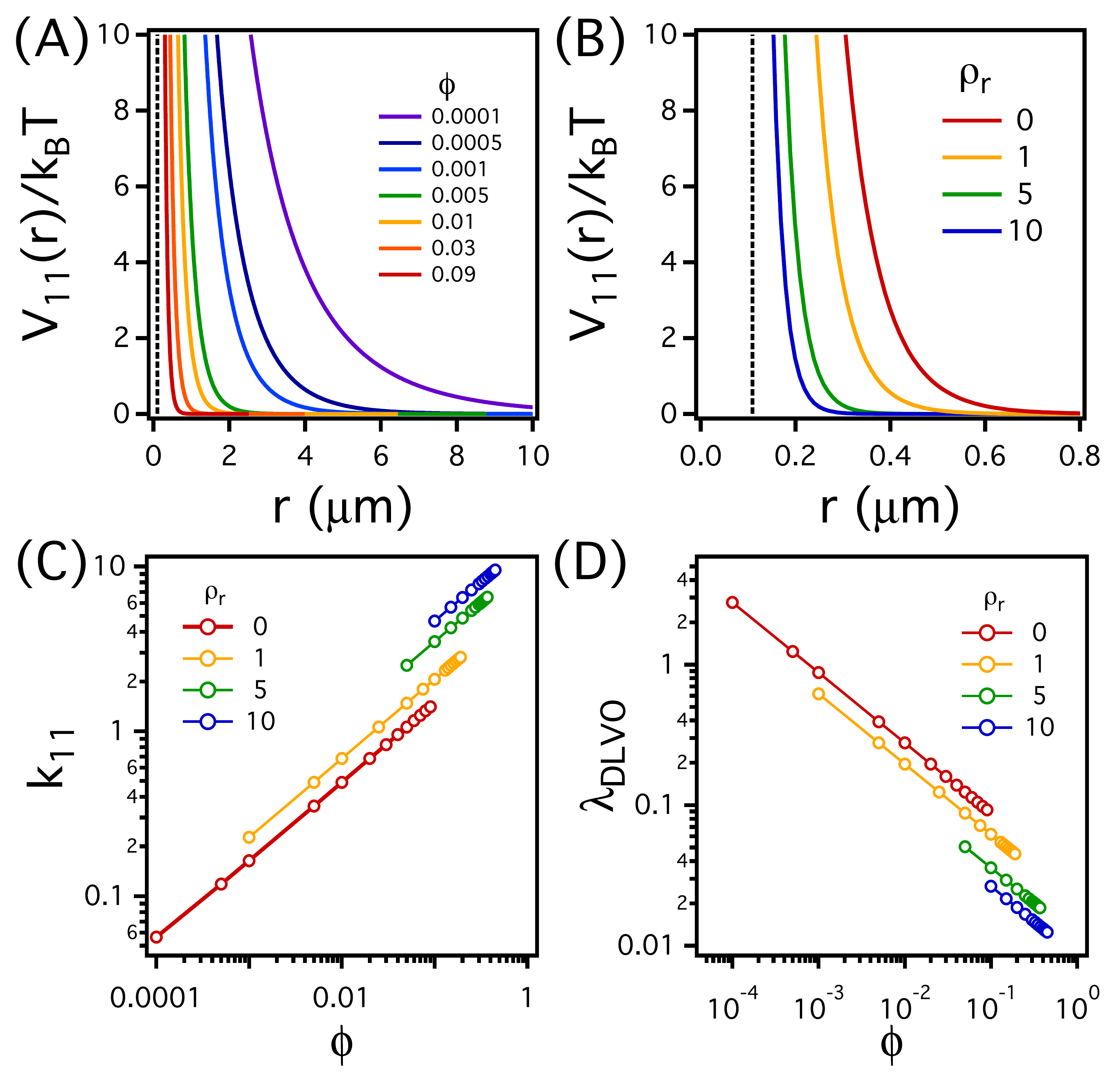}
\caption{\textbf{Variation in the softness of the interparticle potential of the charged colloids as a function of added salt.} Plot of the DLVO potential  (Eq~(\ref{DLVO})) of the small charged colloids $V_{11}(r)/k_BT$ as a function of $r$ (A) for various $\phi$ at $\rho_r=0$, and (B) for various $\rho_r$ with fixed $\phi=0.09$. (C) The stiffness parameter of the small charged colloids $k_{11}$, and (D) the effective length scale of the DLVO potential $\lambda_{DLVO}$ as a function of $\phi$ for various $\rho_r$.}
\label{potential}
\end{figure}

\textbf{Softness of $V_{ij}(r)$:} It is important to note that the \textit{softness} of the interaction of the charged colloids varies with $\rho_r$ and the volume fraction, $\phi=\frac{4\pi}{3L^3}(N_1a_1^3+N_2a_2^3)$. We plot the DLVO potential $V_{11}(r)/k_BT$ for the small colloids as a function of $r$ at different $\phi$ at $\rho_r=0$ (Figure~\ref{potential} (A)) and for different $\rho_r$ at $\phi=0.09$ (Figure~\ref{potential} (B)). The graphs in Figures~\ref{potential} (A) and (B) show that as $\phi$ and $\rho_r$ decrease, the stiffness of the DLVO potential decreases, which means that $V_{11}(r)/k_BT$ decays less steeply as $r$ increases. The effective range of the repulsive interaction increases as $\phi$ and $\rho_r$ decrease. We use steepness and the range of the DLVO potential to characterize the changes illustrated in Figure~\ref{potential} (A) and (B). The stiffness of the DLVO potential can be estimated as a slope of the interaction between colloidal particles as a function of $r$. Since the force is the negative slope of the potential, we define the stiffness parameter $k_{ij}$ as the magnitude of the force of the DLVO potential at $r=l_B$ where $V_{ij}(l_B)=k_BT$. The Debye-H\"{u}ckel screening length $1/\kappa$ is a length scale over which the electrostatic interaction of the charged colloids is effectively screened by other ions. Thus, the effective range $\lambda_{DLVO}$ is $1/\kappa$. In Figures~\ref{potential} (C) and (D), $k_{DLVO}$ and $\lambda_{DLVO}$ are shown for the range of $\phi$ considered in our simulations with varying $\rho_r$, respectively. For a fixed value of $\rho_r$ stiffness $k_{11}$ increases dramatically as $\phi$ increases (Figure~\ref{potential} (D)). Because such changes in the interactions can be readily achieved in experiments, we can use charged colloidal suspensions to investigate how the nature of glass transition itself changes as $\phi$ and $\rho_r$ are varied. 

\section{Results and Discussion}

\subsection{Two transition densities in Wigner glasses}

\begin {figure}
\centering
\includegraphics  [width=3.5in] {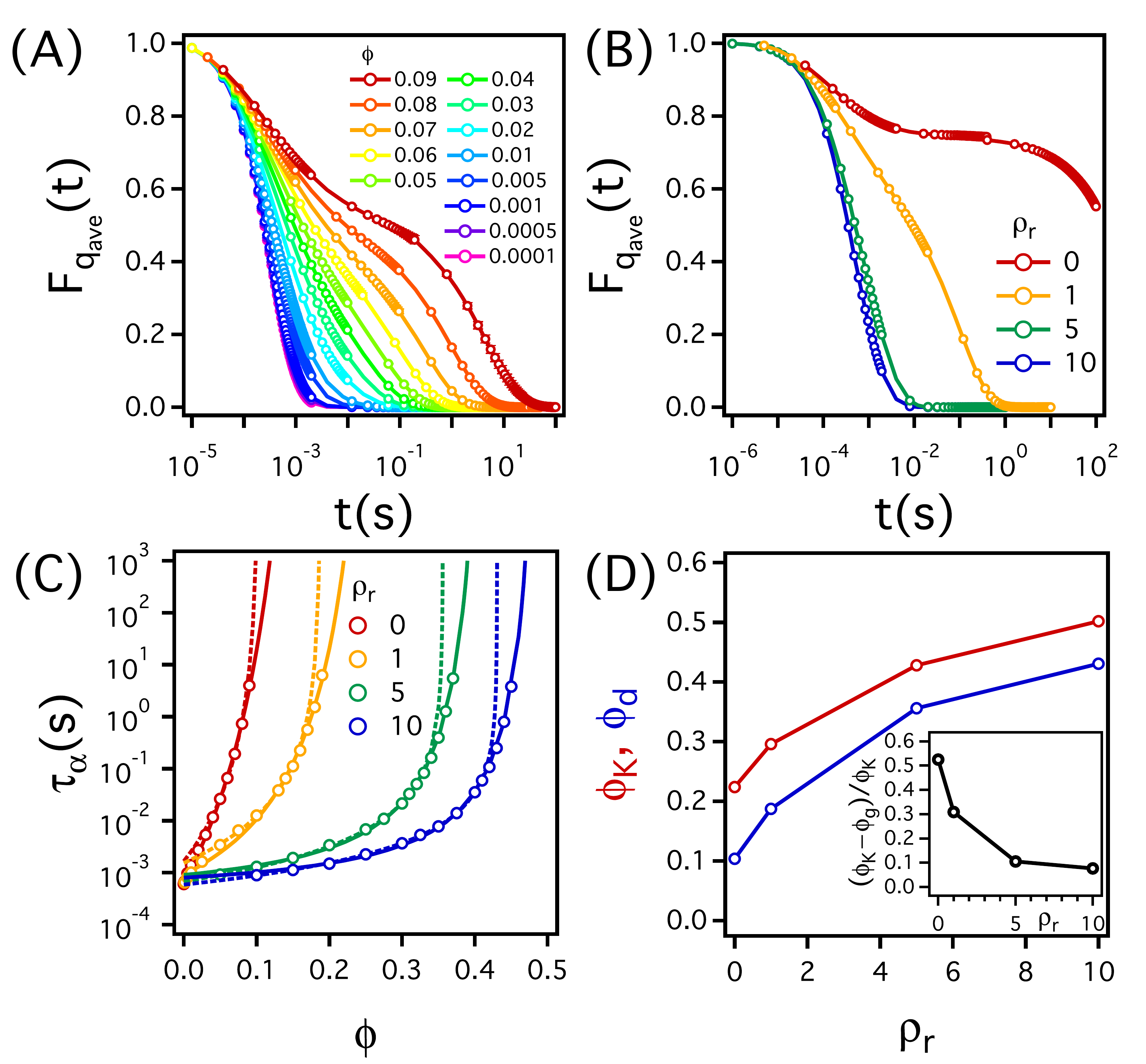}
\caption{\textbf{Structural relaxation and glass transition in charged colloids.} The self-part of the intermediate scattering function $F_{q_{ave}}(t)$ (A) for various $\phi$ with $\rho_r=0$ and (B) for various $\rho_r$ with $\phi=0.15$. (C) The relaxation time of the charged colloids as a function of $\phi$ for various $\rho_r$. The solid and dashed lines indicate VFT and MCT fits to $\tau_\alpha$, respectively.  (D) The characteristic densities $\phi_K$ (red circles) and $\phi_d$ (blue circles) as a function of $\rho_r$.}
\label{Dynamic}
\end{figure}

In order to extract the structural relaxation times as a function of $\phi$ and $\rho_r$, we calculated the self-part of the intermediate scattering function,  
\begin{equation} \label{Fkt}
F_q(t) = \frac{1}{N}\Big\langle\sum_{j=1}^{N}\exp \big [-i\vec{q}\cdot(\vec{r}_j(t)-\vec{r}_j(0)) \big ]\Big\rangle.
\end{equation}   
We used the wavenumber $q=|\vec{q}| = q_{ave}=\frac{2\pi}{d_{ave}}$, where $d_{ave}=2\frac{\phi_1 a_1+\phi_2 a_2}{\phi_1+\phi_2}$ is the volume averaged diameter of the charged colloids with $\phi_1$ and $\phi_2$ being the volume fractions of the small and large colloids, respectively. The reason for using $q_{ave}$ is that that the relaxation times extracted from the time-dependence of $F_{q_{ave}}(t)$ correlates well with  the shear viscosity as a function of  $\phi$ (see Appendix B). In Figure~\ref{Dynamic} (A), we plot the time dependence of $F_{q_{ave}}(t)$ as a function of $\phi$ at $\rho_r=0$. As $\phi$ increases, there is a clear two-step decay in $F_{q_{ave}}(t)$, indicating that the dynamics of the charged colloids becomes glassy as the system is compressed. In Figure~\ref{Dynamic} (B), we show $F_{q_{ave}}(t)$ for various $\rho_r$ at a fixed $\phi=0.15$. As $\rho_r$ decreases, the effective range of the repulsive interaction of the charged colloids increases (Figure~\ref{potential} (B) and (D)), which results in an increase in the effective density of the charged colloids, thus explaining the sluggish dynamics with decreasing $\rho_r$ (Figure~\ref{Dynamic} (B)).

As stated in the Introduction, a key prediction of RFOT is that for a generic glass forming materials that undergo SGT as it is compressed (or supercooled), there are two characteristic transitions. One of them is expected at $\phi=\phi_d$, denoting the start of dynamical arrest. The other is the ideal glass transition at $\phi_K$, which is usually difficult to probe in computer simulations. The onset of sluggish structural relaxation dynamics in Figure~\ref{Dynamic} (A) and (B) can be quantified by an increase in the structural relaxation time, $\tau_\alpha$. For practical purposes, we calculated $\tau_\alpha$ using $t=\tau_\alpha$ at which $F_{q_{ave}}(\tau_\alpha)=0.2$. The dependence of $\tau_\alpha$ as a function of $\phi$ for various $\rho_r$ is shown in Figure~\ref{Dynamic} (C). In the $\phi$ and $\rho_r$ range considered here, $\tau_\alpha$ for a given $\rho_r$, increases by nearly 4 orders of magnitude as $\phi$ increases. 

We analyzed the growth in $\tau_\alpha$ with $\phi$ in terms of the two characteristic transition densities~\cite{Kirkpatrick89PRA,Kirkpatrick15RMP}. As $\phi$ approaches $\phi_d$ an extensive number of metastable glassy states emerge. At densities above $\phi_d$ the system is trapped  in one of the many metastable states for arbitrarily long times, and transport occurs through activated transitions involving crossing growing free energy barriers.    The free energy barrier $\Delta F^\ddagger$ between two adjacent mosaic states scales as the size of the mosaic states $\xi$, and is given by $\Delta F^\ddagger\sim\xi^{d/2}$, where $d$ is the space dimension. The RFOT theory predicts that $\xi$ should increase without bound as $\phi\rightarrow\phi_K$ ($\xi\sim(\phi_K-\phi)^{-2/d}$), which in turn leads to the essential singularity in $\tau_\alpha$ at $\phi_K$. 

The values of $\phi_d$ and $\phi_k$ can be extracted using the data in Figure~\ref{Dynamic} (C). The dynamical transition anticipated in the RFOT theory is consistent with the prediction of the mode-coupling theory (MCT), with the caveat that the power law singularity  is avoided in reality. We calculated $\phi_d$  using $\tau_\alpha\sim(1-\phi/\phi_d)^{-\gamma}$, where $\phi_d$ and $\gamma$ are the fitting parameters. As the density is increased further, $\tau_\alpha$ follows the Vogel-Flucher-Tamman (VFT) relation,  
\begin{equation} \label{vft}
\tau_\alpha = \tau_{VFT}\exp\Big [\frac{D_{VFT}\phi}{\phi_K-\phi}\Big],
\end{equation}  
where $\tau_{VFT}$ is $\tau_\alpha$ for $\phi\rightarrow0$, and $D_{VFT}$ is the fragility parameter. We fit $\tau_\alpha$ to the VFT relation in order to extract $\tau_{VFT}$, $D_{VFT}$ and $\phi_K$. The dashed and solid lines in Figure 2 (C) represent the power laws and the VFT fits. The significant increase in $\tau_\alpha$ is accurately fit  by the  two functional forms in different ranges of $\phi$. 

In Figure 2 (D), we show $\phi_d$ and $\phi_K$ for various $\rho_r$  (see the values in Table 1). Interestingly, as $\rho_r$ increases from 0 to 10, $\phi_d$ and $\phi_K$ increase from 0.10 and 0.22 to 0.46 and 0.50, respectively. Considering that upon an increase in $\rho_r$, the shape of the DLVO potential  becomes more hard sphere-like (Figure~\ref{potential} (A) and (B)), we expect that $\phi_d$ and $\phi_K$ should converge to the values for binary hard spheres ($\phi_d \simeq 0.59$ and $\phi_K \simeq 0.64$ \cite{Brambilla:2009bz}) at high values of $\rho_r$. The results in Figure~\ref{Dynamic} (D) show that $\phi_d$ and $\phi_k$ do increase. However, the numerical values for Wigner glasses and HS differ, which is related to the range of the DLVO potential. We showed sometime ago that pair correlation functions of highly charged spherical colloidal suspensions with bare size $a$ can be mapped onto hard spheres with diameter $d_h$ that is greater than $a$~\cite{Thirumalai:1989il,Rosenberg86PRA}. Thus, a similar mapping would predict that the  volume fractions identified here would be larger if the effective hard sphere diameters are used. With this argument, we conclude that the effective $\phi_K$ for Wigner glasses at high $\rho_r$ would achieve the well-known values for hard sphere systems. The distances $(\phi_K-\phi_g)/\phi_K$ at $\rho_r=10$ (Figure~\ref{Dynamic} (D)) and for hard sphere-like are virtually identical, which is also manifested in the fragile-strong crossover (see below). It is clear that addition of salt influences the softness of the DLVO potential, which drastically alters the glass transition behavior. 

\begin{table*}[ht]
\centering
\begin{tabular}{c c c c c c c c c c c}
\hline\hline
            &  & & \multicolumn{2}{c}{MCT} & &\multicolumn{3}{c}{VFT} & &\\
               \cline{4-5} \cline{7-9}
&$\rho_r$ & & $\phi_d$ & $\gamma$ & & $\tau_{VFT}$ & $\phi_K$ & D & & $\phi_g$\\ [0.5ex] 
\hline
&0& &0.10$\pm$0.03 &4.215$\pm$0.0003& &0.8$\pm$0.1 $\times10^{-3}$ & 0.20$\pm$0.02 & 10$\pm$2 & &0.11\\
&1& &0.187$\pm$0.006 &2.7$\pm$0.3& & 1.7$\pm$0.2 $\times10^{-3}$ & 0.264$\pm$0.006& 3.2$\pm$0.3& &0.20\\
&5& &0.356$\pm$0.003 &1.9$\pm$0.1& &0.9$\pm$0.1 $\times10^{-3}$ & 0.429$\pm$	0.004& 1.4$\pm$0.1& &0.38\\
&10& &0.431$\pm$0.001&1.55$\pm$0.03& &0.79$\pm$0.07 $\times10^{-3}$ & 0.502$\pm$0.004& 0.97$\pm$0.07& & 0.46 \\[1ex]
\hline\hline
\end{tabular}
\caption{Characteristic volume fractions associated with compressed charged colloids}
\label{table:nonlin}
\end{table*}

\subsection{Fragility decreases substantially as $\rho_r$ decreases}

The variations in the fragility can be visualized using the Angell plot, from which the fragility index may be calculated by fitting $\tau_\alpha$ on a logarithmic scale with respect to $\phi/\phi_g$ where $\phi_g$ is the volume fraction at the glass transition. The difference between strong and fragile glasses is evident in the dependence of $\tau_\alpha$, which increases gradually as $\phi/\phi_g$ increases for strong glass formers, but for fragile glass formers the increase is steep. It is worth emphasizing that substantial variations in the curvatures observed in the dependence of $\tau_\alpha$ on $\phi$ (Figure 2(C)) is observed in experiments only when shear viscosity for a very large class of materials as a function of temperature is simultaneously plotted. Remarkably, here in binary mixture of charged suspensions, interacting with isotropic DLVO potential, a similar behavior is observed. The results in Figure 2(C) allows us to construct the Angell plot, $\tau_\alpha$ versus  $\phi/\phi_g$ where $\phi_g$ is the glass transition density.

In order to obtain the Angell plot for Wigner glasses, we first determined $\phi_g$ for various $\rho_r$, which is not a trivial task in the simulations. Typically, $\phi_g$ is obtained from experimental data using $\tau_\alpha(\phi_g)=100s$. For colloidal systems, when $\tau_\alpha=100s$, $\tau_\alpha/\tau_{VFT}\simeq10^5$, thus $\phi_g$ can be obtained using $\tau_\alpha(\phi_g)/\tau_{VFT}=10^5$ \cite{vanderScheer:2017hd,Philippe:2018ee}. In order to calculate $\phi_g$, therefore, one should consider the range of $\phi$ where $\tau_\alpha\simeq 100s$ or $\tau_\alpha/\tau_{VFT}\simeq10^5$, which, unfortunately, is not practical using computer simulations. Alternatively, assuming that $\tau_\alpha$ for high $\phi$ regime follows the VFT relation (Eq~\ref{vft}), we determined $\phi_g$ by extrapolating Eq~\ref{vft} to $\phi$ where $\tau_\alpha/\tau_{VFT}=10^5$ (see Table 1).  The values of $\tau_\alpha$ at $\phi_g$ for various $\rho_r$ in our simulations are $\sim100s$, which justifies our estimation of $\phi_g$. 

\begin {figure}
\centering
\includegraphics  [width=3.5in] {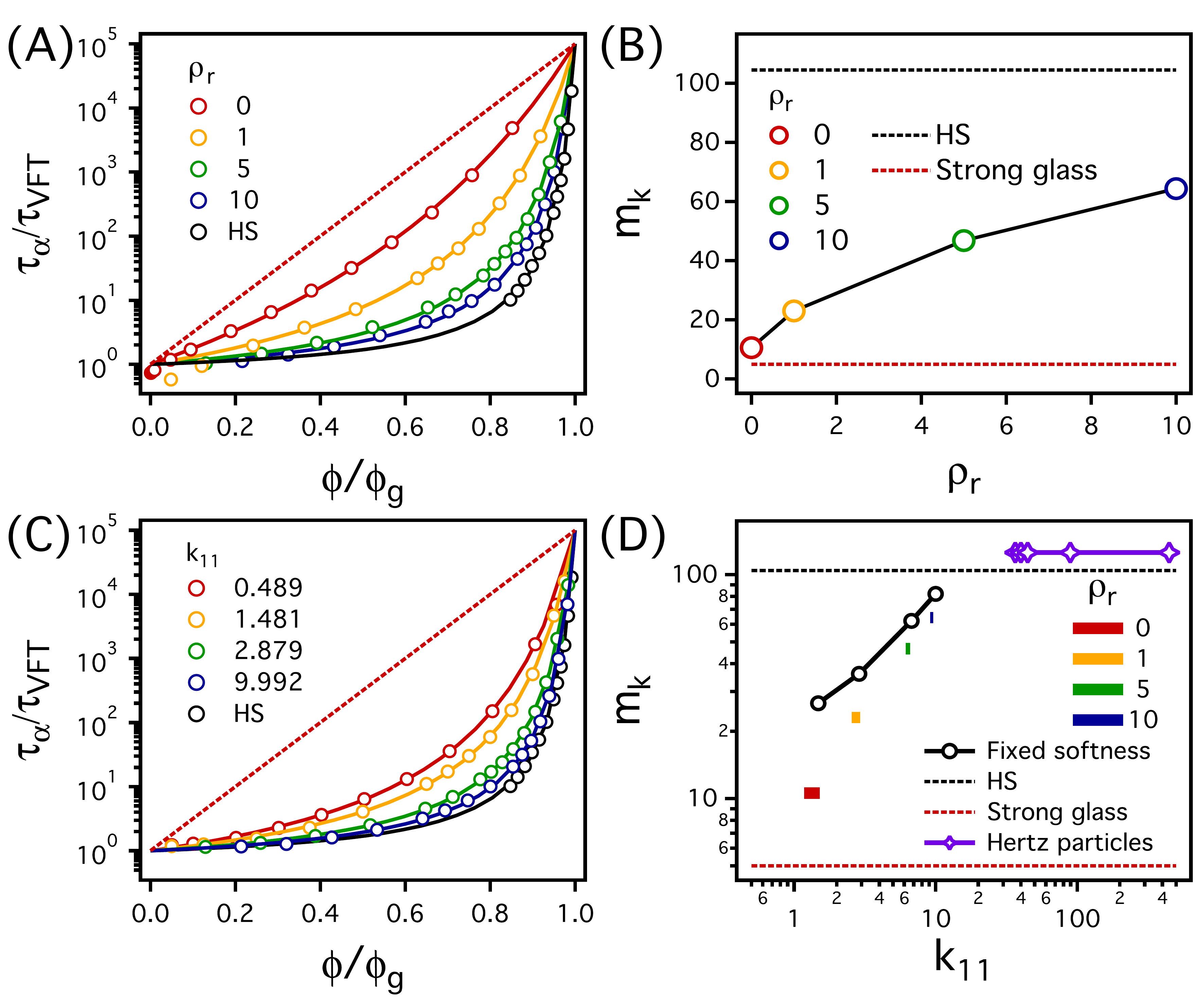}
\caption{\textbf{Variations in the fragility of charged colloids.} (A) Angell plot in which $\tau_\alpha$ for various $\rho_r$ is shown as a function of $\phi/\phi_g$. The black open circles (the rightmost curve) are $\tau_\alpha$ of $\phi$ for hard spheres. The red dashed line shows the Angell plot for a strong glass. The solid lines are the corresponding VFT fits. (B) The fragility index $m_k$ as a function of $\rho_r$. The red and black dashed lines show the fragility of the strong glass and hard spheres, respectively. (C) Relaxation time as a function of as a function of $\phi/\phi_g$ with  the stiffness $k_{11}$ explicitly shown. (D) Fragility index as a function of $k_{11}$. For comparison we also show $m_k$ (Eq. \ref{Frag} for a few other systems.}
\label{VFT}
\end{figure}

The Angell plot for charged colloids for various $\rho_r$ values are shown in Figure~\ref{VFT} (A). We expect $\tau_\alpha$ for strong glasses to increase linearly as a function of  $\phi/\phi_g$~\cite{Mattsson:2009jv,vanderScheer:2017hd}, which implies that $\tau_\alpha\sim\exp [A\phi]$ (the red dashed line in Figure~\ref{VFT} (A)), where $A$ is a constant (the red dashed guide line). On the other hand, for hard sphere colloids that we consider as a reference for a fragile glass (Appendix C), $\tau_\alpha$ increases rapidly as $\phi/\phi_g$ approaches 1 (the black circles). Figure~\ref{VFT} (A) shows that the slope of $\tau_\alpha$ near $\phi/\phi_g=1$ increases significantly as $\rho_r$ increases, indicating that the fragility  increases with $\rho_r$. We quantify the fragility change using the \textit{kinetic fragility index} $m_k$, which is defined as, 
\begin{equation} \label{Frag}
m_k = \frac{d \log \tau_\alpha}{d \phi/\phi_g}\Big |_{\phi=\phi_g}. 
\end{equation}   
For a strong glass, $m_k=5$ since $\log [\tau_\alpha/\tau_{VFT}]$ is linear in $\phi/\phi_g$ (the red dashed line in Figure~\ref{VFT} (B))~\cite{vanderScheer:2017hd}. The value of $m_k$ in the hard sphere limit  is 104 (the black dashed line in Figure~\ref{VFT} (B)). The calculated $m_k$ values of the charged colloids using Eq~(\ref{Frag}), displayed in Figure~\ref{VFT} (B), shows that $m_k$ increases from 10 to 64 as $\rho_r$ increases from 0 to 10. It is likely that if $\rho_r$ is increased further, $m_k$ might converge (perhaps slowly) to the values associated with hard sphere glasses. Therefore, our simulation results demonstrate that by merely tuning the salt concentration, $m_k$ of the charged colloids can be changed dramatically. In other words, addition of a monovalent salt could result in the fragile-strong crossover in Wigner glasses, which is a prediction that can be readily validated experimentally. 

\subsection{Stiffness of the DLVO potential and dramatic fragility changes}

Significant changes in $m_k$ for soft colloids have been reported previously \cite{Mattsson:2009jv,SeekellIII:2015cd,Nigro:2018wb,C7SM00739F}, but the physics underlying this behavior in terms of the  interparticle interaction has remained unclear, and perhaps even controversial. It was found that  there is a crossover from fragile to strong glasses in deformable microgels  when the internal elasticity of the particles decreases. This finding was used to suggest that softer potentials should result in strong glasses~\cite{Mattsson:2009jv}. However, the drastic variations in $m_k$ have not been reproduced in previous simulations by modification of the softness of the potential alone \cite{Ninarello:2017bq,Shi:2011hc,Michele:2004bs,Philippe:2018ee}. For instance, Philippe et al., considered the soft particles interacting via the Hertz potential $V_{H}(r)=\epsilon_H (1-r/\sigma)^2 \Theta (r-\sigma)$, where  $\Theta(x)$ is the Heaviside step function, and $\sigma$ is the diameter of the particles \cite{Philippe:2018ee}. The softness of the potential was tuned by the value of $\epsilon_H$. Although $\epsilon_H$ was varied by two orders of magnitude, the dependence of $\tau_\alpha$ on $\phi$ was found to be insensitive to $\epsilon_H$ (Figure~\ref{VFT} (D)). This implied that the fragility is independent of the softness of the Hertz potential. Thus, they concluded that the idea the softness of the potential solely controls the fragility should be revised. 

Our results for the glass transition of the charged colloids provide insights into these seemingly conflicting arguments. As shown in Figure~\ref{potential} (B), the DLVO potential becomes soft as $\rho_r$ decreases, but even at $\rho_r=0$ the stiffness is altered significantly by $\phi$. These two features of the DLVO potential should contribute to the drastic change in $m_k$ with $\rho_r$ in Figure~\ref{VFT} (B). In order to distinguish between the influence of the softness from that of the other, we carried out additional simulations for the charged colloids whose value of $\kappa$ in Eq~(\ref{kappa}) is fixed as a function of $\phi$. Note that the softness parameter $k_{11}$ and $\lambda$ are determined by $\kappa$. When $\kappa$ is fixed, therefore, the shape (and thereby softness of the potential) is not changed with $\phi$. Accordingly, if the fragility varied with $\kappa$ (or associated softness parameters $k_{11}$ and $\lambda$), this would be attributed solely to the stiffness of the potential. 

Figure~\ref{VFT} (C) shows the Angell plots for various $k_{11}$, from which we evaluate $m_k$ as a function of $k_{11}$ in Figure~\ref{VFT} (D) (the black open circles). They clearly show that as $k_{11}$ decreases (as the potential becomes softer), $m_k$ decreases. In Figure~\ref{VFT} (D) we show $m_k$ for the Hertz particles as a function of $k_{11}$ considered in \cite{Philippe:2018ee} (the purple open stars). The Hertz potential stiffness is expressed as $k_{11}=2\sqrt{\epsilon_H/k_BT}$. In the simulation of \cite{Philippe:2018ee}, the considered range of $\epsilon_H/k_BT$ is from 333 to 50000, which corresponds to  $k_{11}$ in the range from 36 to 447. The $m_k$ values for the Hertz particles is obtained as $m_k\simeq125$ from the simulation data in \cite{Philippe:2018ee}, which is independent of $k_{11}$. The graph shows that the potential in the softest case in \cite{Philippe:2018ee} is steeper than the steepest case in our simulation. This implies that the softness change in \cite{Philippe:2018ee} may not be sufficient to result in a decrease in $m_k$. Therefore, from Figure~\ref{VFT} (D) we conclude that soft potentials (smaller values of $k_{11}$) can indeed make glass transition stronger. 

More importantly, the softness of the potential is  insufficient to fully explain the drastic modification of the fragility shown in Figure~\ref{VFT} (B). We show $m_k$ for various $\rho_r$ in the $y$ axis of Figure~\ref{VFT} (D) (the horizontal color bars) and the $x$ range of each color bar represents the range of $1/k_{11}$ from $\phi_g$ to $\phi=\phi_{ref}$ at which $\tau_\alpha/\tau_{VFT}=10^2$. Note that the $x$ axis is drawn in a log scale, and  thus the range of the color bars in the graph indicates the extent of how sensitively $k_{11}$ changes relatively to $k_{11,\phi_g}$ when $\phi$ drops from $\phi_g$ to $\phi_{ref}$. The graph shows that as $\rho_r$ decreases, $k_{11}$ varies more with $\phi$ and $m_k$ drops further from that of the hard sphere (the black dashed line) comparing to when the softness is fixed with $\phi$ (the black circles). This demonstrates that the variation of the softness with $\phi$ should also play an important role in the drop of the fragility with a decrease in $\rho_r$. 

Note that the potential of the deformable microgels in previous experiments would behave in a similar way upon packing. The microgels deswell significantly upon packing due to their polymeric nature, leading to a decrease in the effective size and an increase in the internal elasticity~\cite{vanderScheer:2017hd,vanderScheer:2017hd}. This may alter the shape of the potential becoming steeper with increasing $\phi$, which is qualitatively similar to the DLVO potential. When deswelling upon packing was limited, the fragility of the migrogels was insensitive to the softness of the potential~\cite{vanderScheer:2017hd}. Therefore, it was suggested recently that rather than softness itself, such variation in the shape of potential with $\phi$ would play a more dominant role in the change in $m_k$ of the microgels \cite{Philippe:2018ee,vanderScheer:2017hd}. 

Figure~\ref{VFT} (D) partly supports this idea. Typical soft microgels in experiments are modeled using the Hertz potential with $\epsilon_H/k_BT\simeq1000$~\cite{C3SM50503K,PhysRevLett.105.025501}. Thus, $k_{11}\simeq60$, which is  steep resulting in the glass transition being  stronger according to Figure~\ref{VFT} (D). This means that the drastic decrease in $m_k$ in microgels could be attributed to other mechanisms associated with their deswelling. However, further investigation is needed  to elucidate how much the potential of the microgels is modified by deswelling and whether this change in the potential  contributes to the fragility of the microgels.   

To conclude this section, we showed that the fragility of the charged colloids decreases drastically as $\rho_r$ decreases. As $\rho_r$ decreases, the DLVO potential becomes softer, changing sensitively with $\phi$, which are the determining factors in the decrease in $m_k$ of the charged colloids. Although our results confirm that the softness of the potential can reduce the fragility it is not a unique contribution to drastic change of the fragility observed in the previous experiments. As a result, the conclusion of the pioneering experiment~\cite{Mattsson:2009jv} should be revisited. \textit{The soft nature of inter particle potential is sufficient but not necessary in order to observe a broad change in the fragility, which is a reflection of the dependence of the relaxation times on $\phi$.}

\

\subsection{RFOT quantitatively accounts for glass transition in charged colloids}

In previous sections, we showed that the nature of glass transition in charged colloids is modified by the addition of monovalent salts. In particular, as $\rho_r$ increases, $\phi_K$ increases and $\tau_\alpha$ increases more steeply as $\phi\rightarrow\phi_K$. In this section, we explore the extent to which the universal aspects of RFOT are manifested in Wigner glasses. We find evidence for strong spatial heterogeneity in the dynamics of the charged colloids as $\phi$ approaches glass transition. We also demonstrate that \textit{dynamic heterogeneity} is closely associated with a significant increase in $\tau_\alpha$, which is a consequence of the increase in length scale as the system is compressed, which was anticipated by the RFOT for the SGT problem. 

\begin {figure*}
\centering
\includegraphics  [width=6.in] {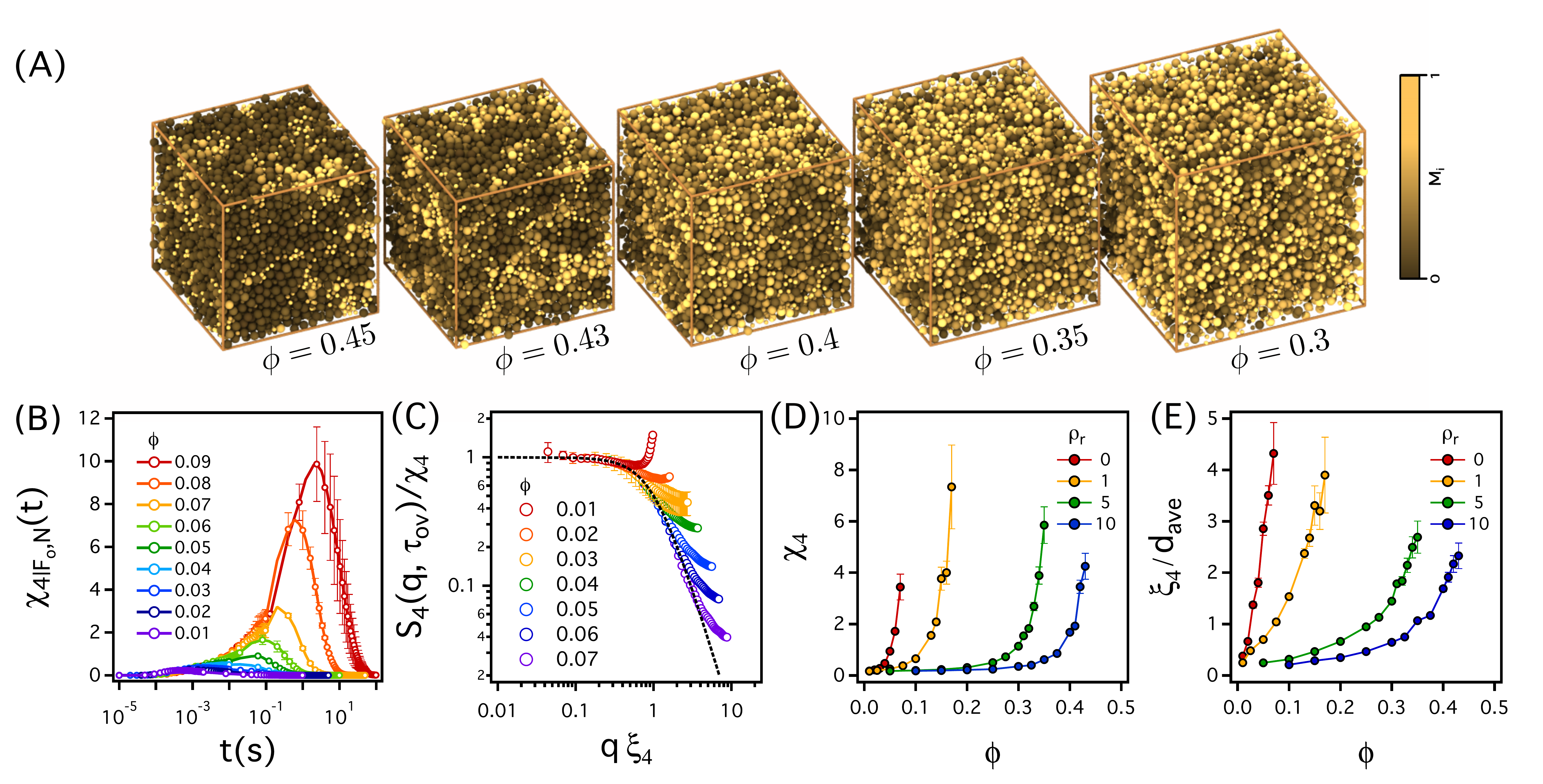}
\caption{\textbf{Visualization and quantification of heterogeneous dynamics in charged colloids.} (A) Snapshots illustrating heterogeneous dynamics. Equilibrium configurations for various $\phi$ with $\rho_r=10$ are prepared.  Coloring of the colloids are done according to their relative mobility $M_i$ (see the definition of $M_i$ in the main text).  (B) The four-point susceptibility of the overlap functions $\chi_{4|F_o,N}(t)$ for various $\phi$ at $\rho_r=0$. (C) The rescaled four-point structure factor $S_4(q,t)/\chi_4$ with respect to the reduced wavenumber $\xi_4$ when $t=\tau_{ov}$ and $\rho_r=0$. The dashed line represents Ornstein-Zernicke fit, $S_4(q,\tau_{ov})/\chi_4=1/[1+(q\xi_4)^2]$. (D) $\chi_4$ and (E) $\xi_4$ of the charged colloids versus $\phi$ for various $\rho_r$.}
\label{hetero}
\end{figure*}

As $\phi$ increases above $\phi_d$, the dynamics of supercooled liquids becomes spatially heterogeneous, which implies that particles with similar mobilities are likely to be localized close to each other. Spatial heterogeneity, resulting in violation of law of large numbers \cite{Thirumalai89PRA,Kang:2013kp,Kirkpatrick15RMP}, is one of the most striking features of the glass transition~\cite{Biroli:2013gy,Ediger:2000ed,Ediger:2012fw}. The heterogeneous dynamics in Wigner glasses is pictorially illustrated in Figure~\ref{hetero} (A). We prepared the simulation snapshots at values of $\phi$ from $\phi=0.3$ to $\phi=0.45$ at $\rho_r=10$. We colored the individual particles according to their relative mobility $M_i$ defined as $M_i=\Delta r_i^2(\tau_{d_{ave}})/\langle\Delta r^2(\tau_{d_{ave}})\rangle$, where $\Delta r_i(t)$ is the displacement of particle $i$ at time $t$, and $\tau_{d_{ave}}$ is a timescale at which the mean-squared displacement $\langle\Delta r^2(\tau_{d_{ave}})\rangle$ is equal to $d_{ave}$. Note that $\tau_{d_{ave}}$ is comparable to the structural relaxation time $\tau_\alpha$, indicating that the structural correlation of whole system would be vanishingly small at $\tau_{d_{ave}}$. A typical liquid is ergodic on the observation time scales $\tau_{obs}$ that is comparable to $\tau_\alpha$. Thus, the structural correlation in  liquids in any large enough subsample would fully vanish for $\tau_\alpha$. As $\phi\rightarrow\phi_d$, however, the system becomes non-ergodic even on $\tau_{obs} > \tau_\alpha$ due to the emergence of an ensemble of disconnected mosaic states. As a consequence, relaxation becomes spatially heterogeneous over large regions. As shown in the figure, as $\phi\rightarrow\phi_d$, the mobilities of the particles are spatially heterogeneous and many of the particles in certain regions rarely diffuse. This reveals that the time evolution of the particles varies from region to region even if  $\tau_{obs}\simeq\tau_{d_{ave}}$. This is a clear indication of the dynamic heterogeneity and is a consequence of broken ergodicity near $\phi_d$, which is an important consequence of the RFOT \cite{Kirkpatrick15RMP}. 

{\bf Fourth order susceptibility:} The extent of dynamic heterogeneity can be quantified by the fluctuations in the two-point correlation function characterizing the structure relaxation. Hence, we should consider the four-point susceptibility, first introduced in \cite{Kirkpatrick:1988bk}, of the intermediate scattering function $\chi_{4|F_{q}} (t) = \frac{1}{N}[\langle F_q(t)^2\rangle-\langle F_q(t)\rangle^2]$. However, the fluctuation of the overlap function is often used as an alternative for numerical convenience, whose behavior is qualitatively similar  to $\chi_{4|F_{q}} (t)$. The overlap function $F_o(t)$ is defined as,  
\begin{equation} \label{overlap}
F_o(t)=\frac{1}{N}\sum_{i=1}^{N}w_i(t), 
\end{equation}   
where $w_i(t) = \Theta(a-|\vec{r}_i(t)-\vec{r}_i(0)|)$ and $\Theta(x)$ is the Heaviside step function, which accounts for the fraction of the slow particles that diffuse a distance  less than $a$. If one uses $a=0.3d_{ave}$, $F_o(t)$ behaves in a qualitatively similar fashion to $F_{q_{ave}}(t)$~\cite{Flenner:2011ht}. Thus, the overlap function can be used to quantify the structural relaxation of liquids instead of $F_{q_{ave}}(t)$. The four-point susceptibility of $F_o(t)$ for the $N$ particle systems is expressed as, 
\begin{equation} \label{overlap_susc}
\chi_{4|F_o,N}(t) =N[\langle F_o(t)^2\rangle-\langle F_o(t)\rangle^2]
\end{equation}  
which characterizes the fluctuation of the slow particles. If the dynamics of liquids becomes spatially heterogeneous, $\chi_{4|F_o,N}(t)$ should have a large value. Accordingly, $\chi_{4|F_o,N}(t)$ can be used as a measure of the overall extent of dynamic heterogeneity. In Figure~\ref{hetero} (B), we plot $\chi_{4|F_o,N}(t)$ as a function of $t$ for various values of $\phi$ at $\rho_r=0$, $a=0.3d_{ave}$ and $N=10,000$. $\chi_{4|F_o,N}(t)$ has a maximum value near $t=\tau_{ov}$, which is the relaxation time of $F_o(t)$ defined as $t$ at which $F_o(t)=0.2$. An increase in $\phi$ results in the maximum value increasing drastically, showing the growth in the extent of dynamic heterogeneity  as $\phi$ approaches the glass transition.

It should be noted that $\chi_{4|F_o,N}(t)$ obtained directly from Eq~(\ref{overlap_susc}) in simulations is subject to strong finite size effects, because the contribution of long-range density fluctuations over the length scale $L$ (the simulation box size) to $\chi_{4|F_o,N}(t)$ is not included \cite{Flenner:2011ht}. A more exact estimation of the extent of dynamic heterogeneity can be achieved by the small wavenumber behavior of the four-point dynamic structure factor $S_4(q,t)$ defined as, 
\begin{equation} \label{FPS}
S_4(q,t) = N[\langle W_o(q,t)W_o(-q,t)\rangle-\langle W_o(q,t) \rangle^2],
\end{equation}  
where $W_o(q,t)=\frac{1}{N}\sum_{j=1}^{N}w_j(t)\exp[-i\vec {q}\cdot \vec{r}_j(0)]$. For small $q$, $S_4(q,t)$ can be fit by the Ornstein-Zernicke equation \cite{Lacevic:2003el,Karmakar:2010fn,Flenner:2010io,Flenner:2011ht},  
\begin{equation} \label{length_eq}
S_4(q,t) \simeq \frac{\chi_{4|F_o,\infty}(t)}{1+(q\xi_{4|F_o}(t))^2} \ \ \text{for} \ q\rightarrow0,
\end{equation}  
where $\chi_{4|F_o,\infty}(t)$ is $\chi_{4|F_o,N}(t)$ of an infinitely large system and $\xi_{4|F_o}(t)$ is a length scale of dynamically correlated regions, both of which can be determined as fitting parameters. The results are shown in Figure~\ref{hetero} (C). The values of $\chi_{4|F_o,N}(t)$ and $\xi_{4|F_o}(t)$ are maximized near $\tau_{ov}$ \cite{Flenner:2011ht}, allowing us to determine the dynamic susceptibility and the associated length scale of the charged colloids as $\chi_4=\chi_{4|F_o,N}(\tau_{ov})$ and $\xi_4=\xi_{4|F_o}(\tau_{ov})$, respectively. Figures~\ref{hetero} (D) and (E) provide a quantitative illustration of the growing dynamic heterogeneity of the charged colloids. They show that $\chi_4$ and $\xi_4$ increase drastically as $\phi$ approaches the glass transition density, implying that a significant slowdown of dynamics near the glass transition is accompanied by heterogeneous dynamics of the charged colloids. 

\begin {figure}
\centering
\includegraphics  [width=3.5in] {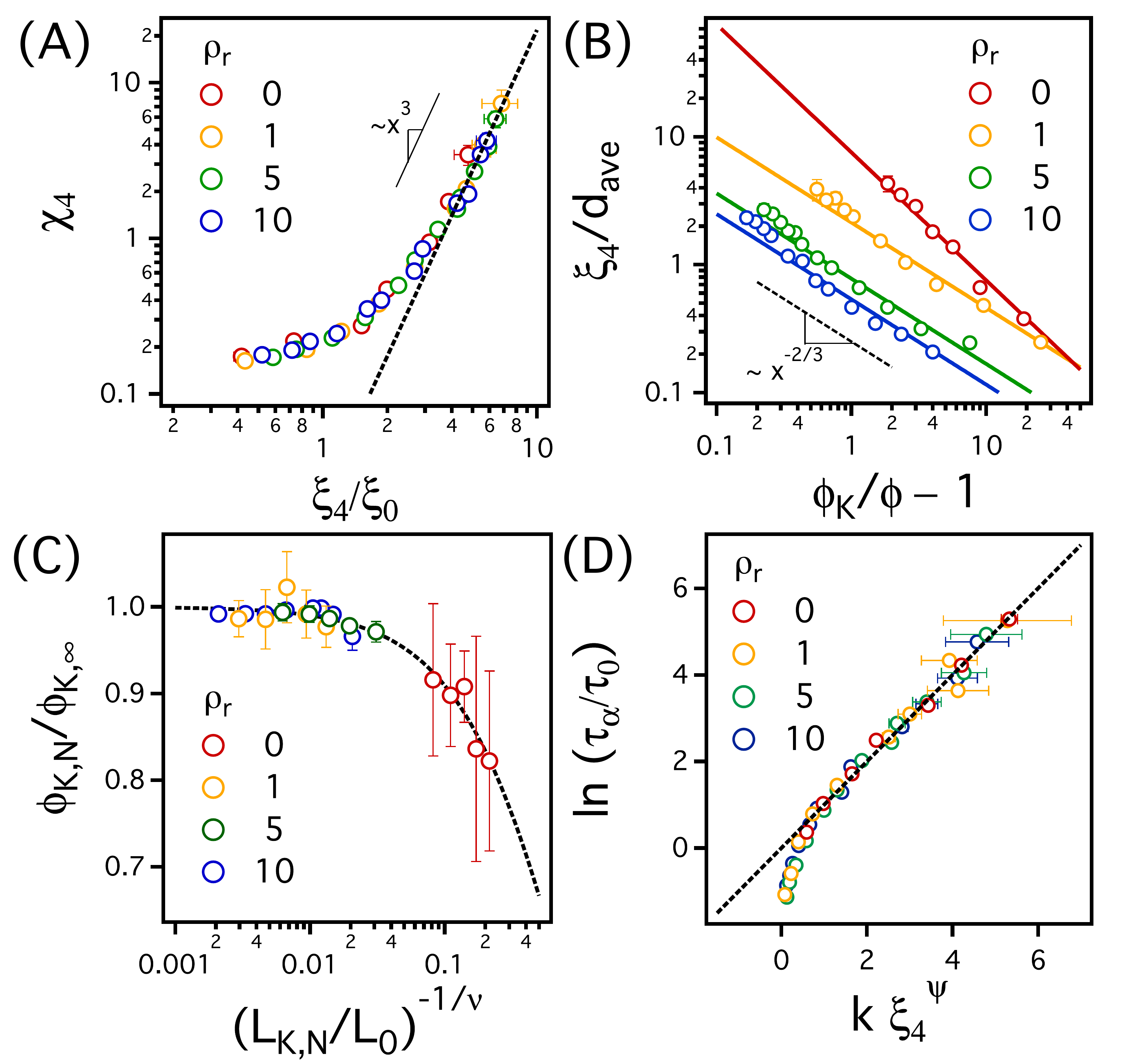}
\caption{\textbf{Illustrating RFOT behavior in  charged colloids.} (A) $\chi_4$ for various $\rho_r$ values as a function of $\xi_4$ extracted using Eq. \ref{length_eq}. The dashed line ($y\sim x^3$) is drawn as a guide to the eye. (B) $\xi_4$ for the various $\rho_r$ as a function of $\phi_K/\phi-1$. For $\rho_r$ the largest $\phi$ for which reliable simulations could be done is less than $\phi_g$ (see Figure~\ref{Dynamic}C), which is a great distance away from the extrapolated $\phi_K$ value (Table 1), which is outside the regime of applicability of the RFOT theory.  The exponent $-2/3$ holds for $\rho_r=$ 1, 5, and 10. (C) Dependence of the reduced ideal glass transition volume fractions $\phi_{K,N}/\phi_{K,\infty}$ on the system size $N$ are shown as $(L_{K,N}/L_0)^{-1/\nu}$ is varied. The dashed line ($\phi_{K,N}/\phi_{K,\infty}-1=(L_{K,N}/L_0)^{-1/\nu}$) validates that the finite size scaling relation predicted by the RFOT theory. Individual fits are shown in Figure~\ref{RFOT_supple} in the Appendix E. (D) The logarithm of the relaxation time $\ln \tau_\alpha$ with respect to $\xi_4^{\psi}$. The dashed guide line ($y=x$) confirms the linear relation between the two quantities. }
\label{RFOT_total}
\end{figure}

{\bf Fractal dimension of dynamically correlated regions:} The RFOT theory provides a comprehensive explanation on how the growth of $\xi_4$ and $\chi_4$ correlates with a sluggish dynamics of the charged colloids. It predicts that as $\phi$ increases beyond $\phi_d$ the nature of the nature of the configurational space partitions into an exponential number of mosaic states. Further compression results in the power law divergence of $\xi$ close to $\phi_K$, which leads to a dramatic increase in $\tau_\alpha$. Since the physical origin of dynamic heterogeneity is closely associated with the emergence of the mosaic states, it is natural to expect that the properties of the dynamically correlated regions characterized by $\xi_4$ and $\chi_4$ to be consistent with those of the mosaic states. This implies that the increase of $\tau_\alpha$ should be described by the growth of $\xi_4$ as the system is compressed. 

Although it is unclear if $\xi_4$ behaves in a qualitatively similar way as $\xi$~\cite{Flenner:2012ev}, many theoretical studies have shown that the growth of $\xi_4$ plausibly captures the important aspects of the RFOT associated with the mosaic states~\cite{Flenner:2015go,Flenner:2011ht,Flenner:2014ck,Flenner:2010io,PhysRevLett.105.055703,Ozawa:2019vd}. For example, according to the RFOT theory, the shape of the mosaic states is string-like but would become compact as $\phi\rightarrow\phi_d$, which can be captured by the change in the shape of the dynamically correlated domains~\cite{Stevenson:2006kf}. Since $\chi_4$ is often interpreted as the number of particles in the dynamically correlated domains~\cite{Berth11Book,Bauer:2013dn}, it follows that if $\chi_4$ is plotted with respect to $\xi_4$ in a log-log scale, the exponent should be associated with the fractal dimension $d_f$ of the dynamically correlated domains. Therefore, if the transition of the shape of the dynamically correlated regions were associated with the mosaic states, $d_f$ should increase to 3 as $\phi\rightarrow\phi_d$, and $\phi$ exceeds $\phi_d$. Flenner et al. computationally showed that the transition in $d_f$ of various types of model glasses occurred universally as $\phi$ (or $T$) approached the dynamic transition points $\phi_d$ (or $T_d$) \cite{Flenner:2014ck}. We observe a similar universal transition in $d_f$  in the charged colloidal glasses. Figure~\ref{RFOT_total} (A) shows $\chi_4$ as a function of $\xi_4$ for various $\rho_r$ values. As shown in the graph, $\chi_4$ vs $\xi_4$ for various $\rho_r$ collapses on to a single curve using an appropriate rescaling factor $\xi_0$, and $d_f$ increase to 3 as $\phi\rightarrow\phi_d$. This particular example confirms the relevance between the dynamically correlated domains and the mosaic states, which guarantees that the growth of $\xi_4$ with $\phi$ characterize the dependence of $\xi$ on $\phi$.  

{\bf Increase in length scale upon compression:} According to the RFOT theory~\cite{Kirkpatrick89PRA,Kirkpatrick15RMP}, $\xi$ should grow as $\phi\rightarrow\phi_K$,
\begin{equation} \label{length}
\xi\sim\Big(\frac{\phi_K}{\phi}-1\Big)^{-\nu},
\end{equation}  
where $\nu = \frac{2}{d}$ is the scaling exponent. We confirm that this is indeed the case when $\xi_4$ is plotted as function of $\phi$. We use $\phi_K$ obtained in Figure~2 (C) and (D), and plot $\xi_4$ for various $\rho_r$ with respect to $\phi_K/\phi-1$ in Figure~\ref{RFOT_total} (B). The slopes of the linear guide lines in the graph represent the values of $\nu$ which are expected to be $2/d$ by the RFOT theory. For $\rho_r$ =1, 5, and 10, $\xi_4$ of $\phi$ follows well Eq~\ref{length} with $\nu=2/3$. For completeness, we also plot the results for $\rho_r=0$.   In principle, Eq~(\ref{length}) is applicable only if $\phi$ is relatively close to $\phi_K$. For $\rho_r=0$, the distances ($\Delta \phi_K = 1-\phi/\phi_K$) between $\phi_K$ and the maximum $\phi_{max}$ for which reliable simulations can be performed (Figure~\ref{RFOT_total} (B)) is 0.65, which is two to four times larger than for $\phi_r=1$, 5, and 10. This results in the deviation from the expected  value in $\nu$ when $\rho_r=0$. Thus, to more precisely determine $\nu$ for $\rho_r=0$, it is necessary to estimate $\xi_4$ at larger values of $\phi$, which is difficult to do in simulations. Nevertheless, the results in Figure~\ref{RFOT_total} (B) are sufficient to demonstrate that, as predicted by RFOT, $\xi_4$ for various $\rho_r$ of the charged colloids would diverge at $\phi_K$, thus validating one of the key predictions of the RFOT theory.  

In order to show that the divergence of $\tau_\alpha$ at $\phi_K$ is due to the increase in $\xi\simeq\xi_4$, we investigated the finite size effects on the estimates of $\phi_K$. Let us assume that the system with size $L$ is a subsample of an infinitely large system whose $\phi_K$ is $\phi_{K,\infty}$. 
Then, $\phi_K$ of $L$ would be determined as $\phi$ at which $\xi_4\simeq L$. 
Therefore, $\phi_{K}$ and $L$ of the subsystems would follow Eq~\ref{length} with respect to $\phi_{K,\infty}$, which leads to the finite size scaling relation of $\phi_K$, 
\begin{equation} \label{FSS}
L\sim\Big(\frac{\phi_{K,\infty}}{\phi_{K,L}}-1\Big)^{-\nu},
\end{equation}  
where $\phi_{K,L}$ is $\phi_K$ of the subsystem of size $L$. When $N$ is fixed, as in our simulation, $L$ is a function of $\phi$, $L_N\sim(N\phi)^{-1/3}$. Hence, for constant $N$ system, $\phi_{K,N}$ would coincide with  $\phi$, which is determined using $\xi\simeq L_{N,K}\sim(N\phi_{K,N})^{-1/3}$. This leads to $L_{K,N}\sim(\phi_{K,\infty}/\phi_{K,L}-1)^{-\nu}$, from which $\phi_{K,N}$ can be expressed as a function of $L_{K,N}$ as below,
\begin{equation} \label{FSS1}
\phi_{K,N}=\frac{\phi_{K,\infty}}{1+\big(\frac{L_{K,N}}{L_0}\big)^{-1/\nu}},
\end{equation}
where $\phi_{K,N}$ indicates $\phi_K$ associated with finite $N$ and we define $L_{K,N}$ as $L_{K,N}=(N/\phi_{K,N})^{1/3}$. We estimate $\phi_{K,N}$ at various $N$ using the VFT relation as done in Figure~\ref{Dynamic} (C) and determine $L_0$ and $\phi_{K,\infty}$ as fitting parameters in Eq~\ref{FSS1}. For the purpose of fitting the curves, $\nu=2/3$ is used for $\rho_r=1,5,$ and 10, but we use an effective value $\nu=1$ for $\rho_r=0$. Figure~\ref{RFOT_total} (C) shows that $\phi_{K,N}/\phi_{K,\infty}$ for various $\rho_r$ collapses onto a single curve ($y=(1+x)^{-1}$) when plotted against $L_{K,N}^{-1/\nu}$, indicating that $\phi_{K,N}$ follows the scaling relation in Eq~\ref{FSS1}. This clearly demonstrates that it is the growth of $\xi_4$ that contributes to the slow dynamics of the charged colloids as $\phi_K$ is approached.

{\bf Free energy barrier and growing length scale:} The quantitative relation between the growing length scale associated with dynamic heterogeneity and the slowing down of the structural relaxation in Wigner glasses can be explained using RFOT theory \cite{Kirkpatrick89PRA}. The activation free energy $\Delta F^\ddagger$ of a configuration to move from one mosaic state to another increases with an increase in the length scale associated with the mosaic states, i.e., $\Delta F^\ddagger\sim \xi^{\psi}$, where $\psi$ is a scaling exponent. This implies that the structural relaxation time is related to $\xi$ as, 
\begin{equation} \label{RFOT}
\tau_\alpha=\tau_0\exp[k\xi^{\psi}], 
\end{equation}  
where $\tau_0$ is a timescale at $\xi\rightarrow0$ which is comparable to $\tau_{VFT}$ and $k$ is a prefactor for $\Delta F^\ddagger$. Note that $\psi$ is predicted to be $1/\nu$, such that by substituting $\xi$ with Eq~(\ref{length}), Eq~(\ref{RFOT}) can recover the VFT relation. We determine $\tau_0$ and $k$ at various values of $\rho_r$ by fitting $\tau_\alpha$ to Eq~(\ref{RFOT}). Figure~\ref{RFOT_total} (D) shows $\ln [\tau_\alpha/\tau_0]$ at various $\rho_r$ values plotted against $k\xi_4^\psi$, where $\psi=1/\nu$ in Figure~\ref{RFOT_total} (B). The linear line ($y=x$) in the graph is drawn as a guide to confirm the quantitative consistency between the two quantities. This figure clearly illustrates that as $\phi\rightarrow\phi_K$, $\tau_\alpha$ for Wigner glasses at various $\rho_r$ values can be quantitatively described by Eq~(\ref{RFOT}).  

\subsection{Analogies between Wigner and molecular glasses}

\begin {figure}
\centering
\includegraphics  [width=3in] {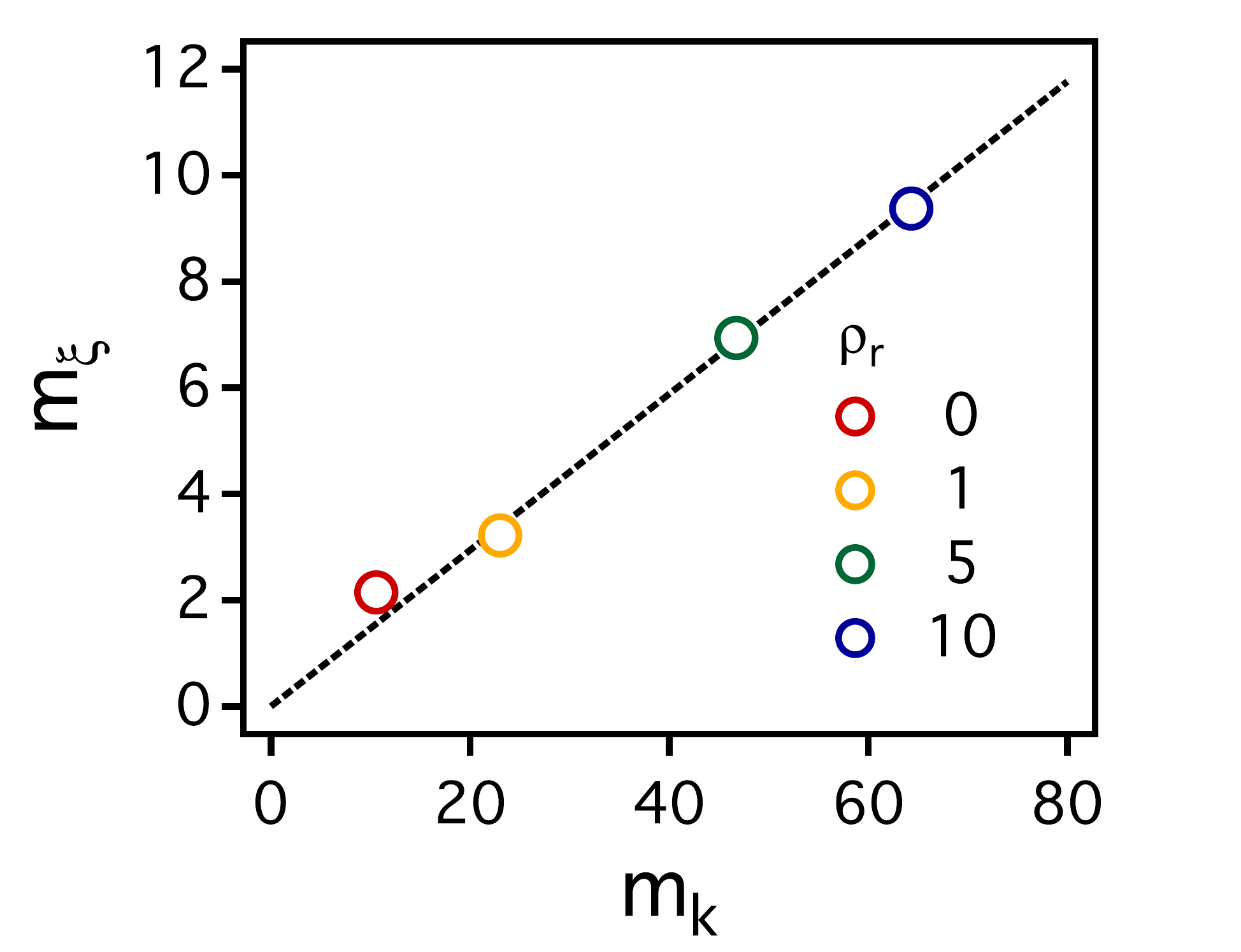}
\caption{\textbf{The linear relation between $m_\xi$ and $m_k$.} $m_\xi$ for the charged colloids as a function of $m_k$ for various $\rho_r$. The dashed line in the graph confirms a linear relation between two quantities.}
\label{Frag_Xi}
\end{figure}

We investigated how the growing length scale relates to the fragility of the charged colloids. By replacing $\tau_\alpha$ in Eq~(\ref{Frag}) with Eq~(\ref{RFOT}), we can define the fragility index $m_\xi$ in terms of $\xi$, i.e., 
\begin{equation} \label{Frag_xi}
m_\xi=k\frac{d \xi^\psi}{d \phi/\phi_g}\Big |_{\phi=\phi_g}, 
\end{equation} 
where $\xi_{\phi_g}$ is $\xi$ at $\phi=\phi_g$. Since $k\xi_{\phi_g}^\psi=\ln [\tau_\alpha (\phi_g)/\tau_0]$ and $\tau_\alpha (\phi_g)/\tau_0\simeq 10^5$ for the colloidal glass, $k\xi_{\phi_g}^\psi$ should have a trivial value. Thus, $m_\xi$ can be expressed as, 
\begin{equation} \label{Frag_xi_reduce}
m_\xi=\psi k\xi_{\phi_g}^{\psi-1}\frac{d \xi}{d \phi/\phi_g}\Big |_{\phi=\phi_g}\sim \frac{d \ln \xi}{d \phi/\phi_g}\Big |_{\phi=\phi_g}. 
\end{equation}
We evaluated $m_\xi$ using the last term in Eq~(\ref{Frag_xi_reduce}) by extrapolating Eq~(\ref{RFOT}) to $\xi_4=\xi_{4,\phi_g}$. In Figure~(\ref{Frag_Xi}), we compare $m_\xi$ with $m_k$, which shows that the two quantities are linearly related.

It is important to note that since the configurational entropy $S_c$ decreases as $\phi\rightarrow\phi_K$ (or $T\rightarrow T_K$) \cite{Ozawa:2018im,Berthier:2017du,Berthconfigu}, i.e., $S_c\sim \phi_K/\phi-1$ (or $S_c\sim T/T_K-1$), $m_\xi$ can be written as a function of $S_c$, 
\begin{equation} \label{Frag_Sc}
m_\xi\sim\frac{d S_c^{-1}}{d \phi/\phi_g}\Big |_{\phi=\phi_g},
\end{equation} 
which is consistent with the definition of the \textit{thermodynamic fragility index} of liquids \cite{Martinez:2001bi}. Therefore, the linear relation between $m_\xi$ and $m_k$ indicates that the thermodynamic and kinetic fragility indices of the charged colloids are proportional to one another. Such a proportionality has also been found for various types of molecular glasses. Martinez and Angell measured the kinetic and thermodynamic fragility index of various organic glasses and found that they had a linear relation for both strong and  fragile glasses \cite{Martinez:2001bi}. Similar behavior was also found by theories and simulations in polymeric glasses \cite{Dudowicz:2005eg,Starr:2011fn}, which implies that glass transition in soft colloids exhibits strong analogies to other molecular glasses. This also shows that the glass transition in simple and complex liquids manifest the universal features predicted by the RFOT theory. 

\section{Conclusions}

We investigated the glass transition in binary charged charged colloids, which are excellent experimentally controllable model soft glasses, by performing extensive Brownian dynamics simulations.  We showed that as concentration of monovalent salts is increased, the inter particle potential becomes soft. As a result,  the characteristic glass transition volume fractions $\phi_d$ and $\phi_K$ decreases. In addition, we predict that  by simply tuning the salt concentration, the fragility of the charged colloids  varies greatly.  The fragility value changes from  about 10 at low salt concentration to in excess of 50 at high salt concentration values. Typically, the large values are associated with molecular systems that interact via anisotropic potentials.   Surprisingly, in Wigner glasses, with isotropic inter particle interactions, the fragility varies continuously by altering a single from a low to high value by tuning a single externally controllable parameter.  

Despite the changes in the softness of potential that  is altered by addition of monovalent salts, the sluggish dynamics can be quantitatively described in terms of the enhanced dynamic heterogeneity as predicted by the RFOT theory. With the four-point dynamic structure factor as an appropriate order parameter, we determined the dynamic susceptibility and the associated increase in length scale by compressing the system.  The simulations unambiguously show that the growth in the length scale with increasing $\phi$ is closely associated with the glassy dynamics, which is strikingly consistent with the prediction of the RFOT theory. 

Using the  RFOT theory, we found that the kinetic fragility index of the colloid can be expressed in terms of the length scales of the heterogeneous dynamics. This indicates that the kinetic and the thermodynamic fragility indices of the charged colloids should be linearly related as has been noted for both molecular glasses, such as organic and polymeric glasses. Thus,  glass transition of diverse materials should be governed by the same universal principles, as anticipated by RFOT, regardless of a broad spectrum of their fragilities.  The present simulations and recent developments show that RFOT provides a comprehensive theory of the structural glass transition by capturing quantitatively the onset of non-ergodicity, and divergence of a growing correlation length as the ideal transition density (or temperature) is approached from above. Because in Wigner glasses there is a smooth crossover from fragile to strong glass behavior, we can conclude that the single unified RFOT is sufficient to describe almost all aspects of glass forming materials. The results in this work firmly establishes that RFOT is likely to be the theory for the structural glass transition.

\

\section{Acknowledgements}
This work was supported in part by a grant from National Science Foundation (CHE 19-00093) and the Collie-Welch Chair (F-0019) administered through the Welch Foundation.

{\bf Appendix A: Brownian Dynamics simulations}

The position of the charged colloids is evolved using Brownian dynamics (BD) by numerically integrating the following equations of motion,  
\begin{equation} \label{BD}
\frac{d\vec{r}_i}{dt} = \frac{D_{i,0}\vec{F}_i(t)}{k_BT} \delta t + \sqrt{2D_{i,0}}\vec{R}_i(t).
\end{equation}
In Eq.\ref{BD}, $\delta t$ is the integration time step, $\vec{r}_i(t)$ is the position vector of the $i^{th}$ particle at time $t$, $\vec{F}_i(t)$ is the total systematic force acting on particle $i$, and $\vec{R}_i$ is the term for the fluctuation force that  satisfies $\langle\vec{R}_i(t)\rangle=0$ and $\langle\vec{R}_i(t)\cdot\vec{R}_i(t')\rangle=6D_{i,0}\delta_{ij}\delta(t-t')$, where $\delta_{ij}$ and $\delta(t-t')$ are the Kronecker $\delta$ and the Dirac $\delta$ function, respectively. The diffusion coefficient, $D_{i,0}$, of the  $i^{th}$ particle in the infinitely dilute regime  is chosen as 4.53 $\mu m^2/s$ and 2.24 $\mu m^2/s$ for the small and large colloids, respectively. Various values of the integration time step $\delta t$ are considered for accuracy and efficiency of the simulation. We use $\delta t = 1 \ \mu s$ for $\rho_r=5$ and  $10$,  $\delta t = 5 \ \mu s$ for $\rho_r=1$ and  $\delta t = 10 \ \mu s$ for $\rho_r=0$, which provides numerical accuracy. 

In order to generate equilibrium configurations, first we randomly placed the binary colloids in the cubic simulation box with periodic boundary conditions in all directions. The simulation box size $L$ is determined by the value of $\phi$, ranging from 4.1 $\mu m$ to 67.9 $\mu m$. Then, we performed simulations for $t=10\tau_\alpha\sim100\tau_\alpha$ to obtain the equilibrium configuration, where $\tau_\alpha$ is the structural relaxation time.  After the equilibration step, we carried out additional BD simulations to calculate various dynamic properties of the system. We consider three independent trajectories for each set of $\phi$ and $\rho_r$ to estimate ensemble averages of various properties. All the simulations were carried out with LAMMPS. 

{\bf Appendix B: Comparison of structural relaxation time with viscosity}

The self-part of the intermediate scattering function $F_q(t)$ characterizes the time-dependent density fluctuation on the length scale $l_q=2\pi/q$. With appropriate choice of $q$, the relaxation time of $F_q(t)$ could capture the viscosity change of liquid near the glass transition~\cite{Sengupta:2013dg}. Although the equivalence between $\tau_\alpha$ and the shear viscosity $\eta$ is often made, it is not always valid. Nevertheless, $F_q(t)$ is commonly estimated both in simulations and experiments to investigate the dynamics of the SGT.

\begin {figure}
\centering
\includegraphics  [width=3.in] {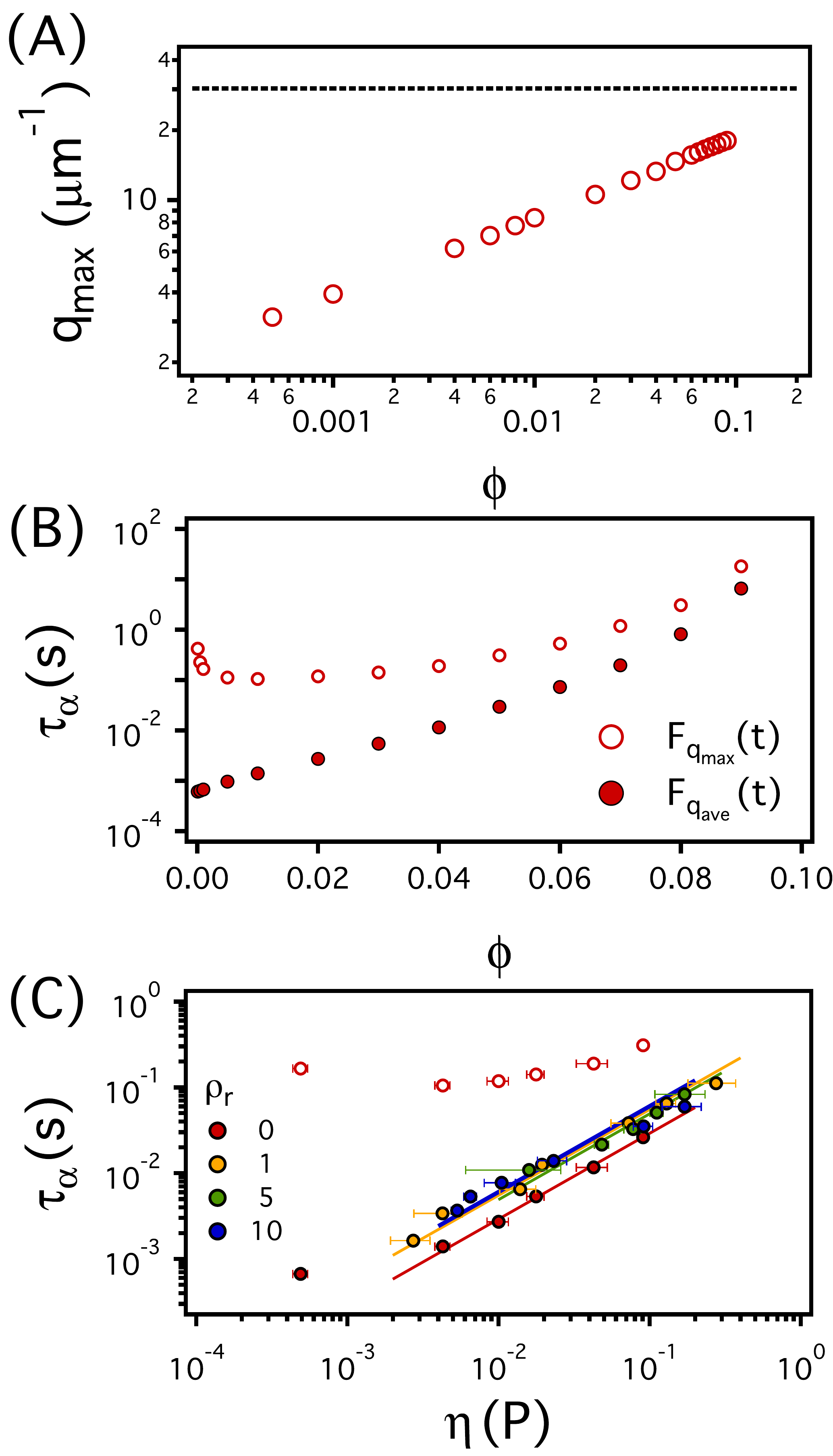}
\caption{ (A) Dependence of $q_{max}$ as a function of $\phi$ when $\rho_r=0$ (the red circles); $q_{max}$ is given by $2 \pi / l_{max}$, where $l_{max}$ is a length scale at which the radial distribution function has the first maximum.  The black dashed line shows that $q_{ave}=2\pi/d_{ave}$, where $d_{ave}$ is the weight-averaged diameter of the charged colloidal particles, is independent of $\phi$. (B) Comparison of $\tau_\alpha$ obtained by $F_{q_{max}}(t)$ (the open red circles) and $F_{q_{ave}}(t)$ (the filled red circles) when $\rho_r=0$. (C) Comparison of $\tau_\alpha$ with $\eta$ for various $\rho_r$. $\tau_\alpha$ of the colored filled circles is obtained by $F_{q_{ave}}(t)$. The colored guide lines ($y\sim x$) confirm the linear relation of $\eta$ with $\tau_\alpha$ from $F_{q_{ave}}(t)$. The open red circles represent $\tau_\alpha$ from $F_{q_{max}}(t)$ when $\rho_r=0$, clearly indicating that $F_{q_{max}}(t)$ fails to capture the changes in $\eta$. }
\label{Viscos}
\end{figure}

Typically, two types of $q$ are used,. One is $q_{max}$ corresponding to the length scale $l_{max}$ at the first maximum of the radial distribution function, and the second is $q_{ave}$ related to the particle size. For liquids with steep potential such as hard spheres, $q_{max}$ is comparable to $q_{ave}$, and is only marginally changed with a change in $\phi$. Thus, the relaxation time $\tau_\alpha$ obtained using the time-dependence of the $F_{q_{max}}(t)$ and $F_{q_{ave}}(t)$ does not result in qualitative difference in the characterizing relaxation dynamics in supercooled liquids. As shown in Figure~\ref{Viscos} (A), however, for charged colloids, $q_{max}$ varies significantly as $\phi$ increases although $q_{ave}$ is invariant. This leads to a qualitatively different behavior of $\tau_\alpha$ extracted from $F_{q_{max}}(t)$ and $F_{q_{ave}}(t)$ upon an increase in $\phi$. When $\rho_r=0$, for example, $\tau_\alpha$ obtained from the decay of $F_{q_{ave}}(t)$ increases monotonically with $\phi$, whereas that calculated from $F_{q_{max}}(t)$ has a minimum value (Figure~\ref{Viscos} (B)). This implies that the description of glass transition of the charged colloids should depend on the value of $q$.

We confirm that $\tau_\alpha$ extracted from the decay of $F_{q_{ave}}(t)$ is more relevant to the viscosity change as the system is compressed. We evaluate the shear viscosity $\eta$ of the charged colloid using Green-Kubo formula, 
\begin{eqnarray}\label{viscosity}
\eta &=& \frac{V}{3k_BT}\int_0^\infty dt \sum_{(\alpha,\beta)^{'}}   \langle P_{\alpha,\beta}(t) P_{\alpha,\beta}(0) \rangle,
\end{eqnarray}
where $\alpha$ and $\beta$ denote Cartesian components ($x$, $y$ and $z$), $(\alpha,\beta){'}$ indicates the sum is over three different combinations of $\alpha$ and $\beta$. The pressure tensor $P_{\alpha,\beta}$ is defined as, 
\begin{eqnarray}\label{pressure}
P_{\alpha,\beta} &=& \frac{1}{V}\sum_{i>j}\frac{r_{ij,\alpha}r_{ij,\beta}}{r_{ij}}\frac{\partial V(r_{ij})}{\partial r_{ij}}.
\end{eqnarray}
In Eq~(\ref{pressure}), $r_{ij}$ is the distance between particles $i$ and $j$. The subscript $\alpha$ under the variable represents the $\alpha$ component in Cartesian coordinate. In Figure~\ref{Viscos} (C), we plot $\tau_\alpha$ versus $\eta$ for various $\rho_r$. The colored filled circles show $\tau_\alpha$ extracted from the time dependent behavior of$F_{q_{ave}}(t)$ as a function of $\eta$. The linear lines ($y\sim x$), used as a guide to the eyes, confirm the linearity between $\tau_\alpha$ and $\eta$, indicating that $\tau_\alpha$ from $F_{q_{ave}}(t)$ captures the viscosity change of the charged colloidal particles accurately. On the other hand, the open red circles represent $\tau_\alpha$ from $F_{q_{max}(t)}$ when $\rho_r=0$, show that $\tau_\alpha$  calculated from $F_{q_{max}}(t)$ and $\eta$ do not correlate. Therefore, we use $F_{q_{ave}}(t)$ to characterize the glassy dynamics  of the charged colloids.  

{\bf Appendix C: Hard sphere system as a reference for fragile glasses}

To consider a reference system for fragile glasses, we perform a dynamic Monte Carlo simulation using binary hard sphere mixtures, which is found to reproduce well the dynamic light scattering (DLS) experiments of poly(methyl methacrylate) (PMMA) particles~\cite{Brambilla:2009bz}. Diameters of the big and small hard spheres are $D_b = 1.4\sigma$ and $D_s=1.0\sigma$, respectively, where $\sigma$ is the reduced unit of length. We consider $N=1000$ hard spheres (the numbers of big and small hard spheres are equal). The positions of the hard spheres are evolved using dynamic Monte Carlo (dMC) simulation with a standard Metropolis algorithm. At every Monte Carlo step, a particle is randomly chosen and displaced by a random vector, whose components for each direction are randomly drawn between $-0.1\sigma$ and $0.1\sigma$. The unit time $t$ is defined as the number of MC steps divided by $N$. Initial configuration is prepared by randomly placing the non-overlapped hard spheres in a periodically replicated (in all dimensions) three dimensional simulation box.   We equilibrated the initial configurations by performing dMC simulations for $t=10\tau_\alpha\sim100\tau_\alpha$, then production run was performed to obtain the relaxation time $\tau_\alpha$ which is determined as a characteristic time scale in the decay of  $F_q(t)$ with $q\sigma=6.1$. We considered $\phi$ from 0.5 to 0.59. For each volume fraction $\phi$, three to five ensembles are used to perform an ensemble average.   

{\bf Appendix D: Peak values of the radial distribution function as a function of $\phi$}

For hard sphere colloids, the height of the peak of the static structure factor $S(q)$ or that of the radial distribution function $g(r)$ increases monotonically with an increase in $\phi$. For the soft colloids, however, they first increase with $\phi$ but begin to decrease at a certain volume fraction~\cite{Zhang:2009bv,C2SM27654B}. This is attributed to non-equilibrium dynamic behavior due to aging or a compressed exponential decay of $F_q(t)$~\cite{Gnan:2019ek,Philippe:2018ee}.

\begin {figure}
\centering
\includegraphics  [width=3.in] {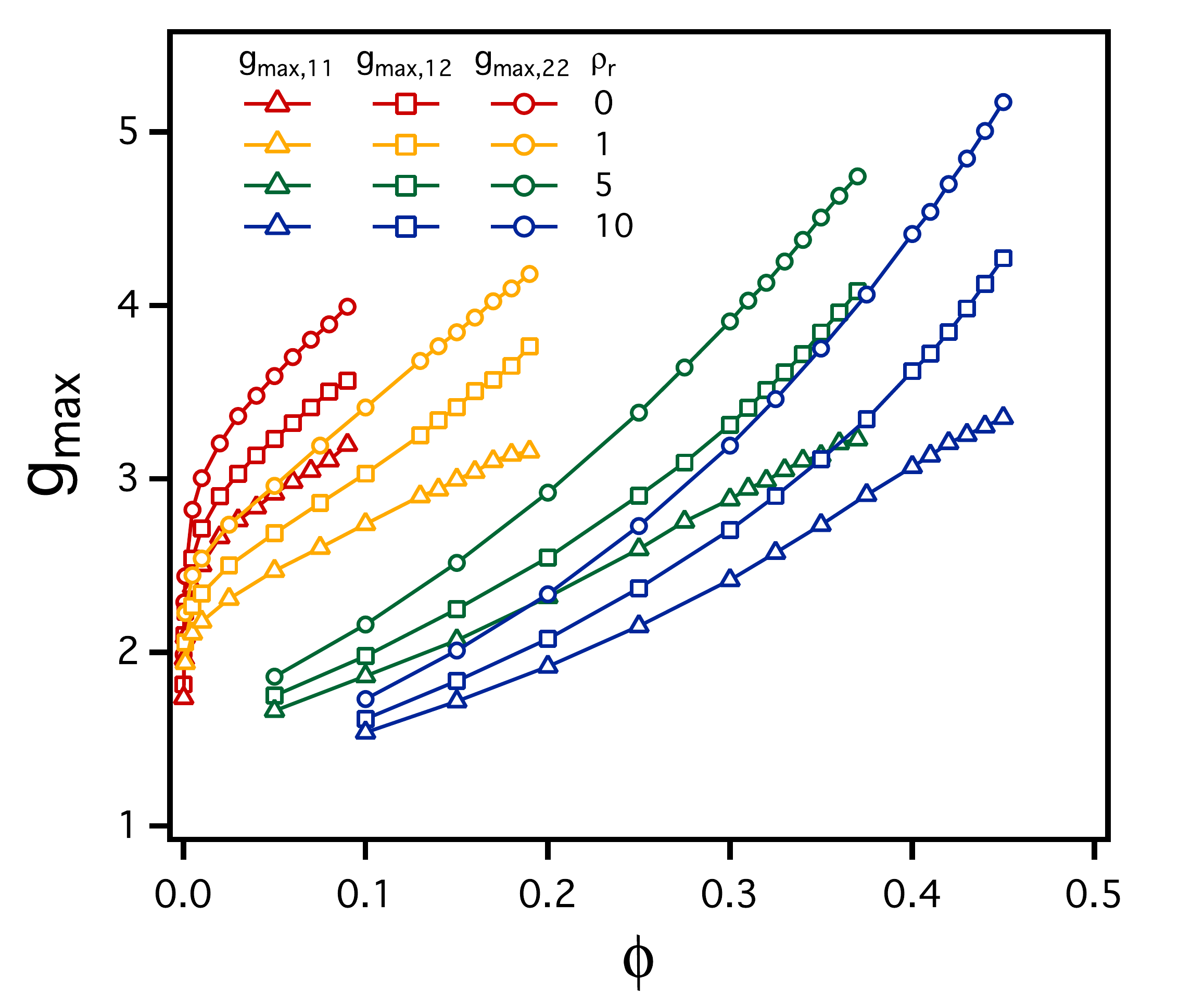}
\caption{$g_{max,11}$, $g_{max,12}$ and $g_{max,22}$ of the charged colloids as a function of $\phi$ at various $\rho_r$.}
\label{grmax}
\end{figure}

In the range of $\phi$ considered in this work, we did not such non-equilibrium effects. We consider three radial distribution functions $g_{11}(r)$, $g_{12}(r)$, and $g_{22}(r)$ of the charged colloids, where $g_{i,j}(r)$ is the radial distribution function between $i$ and $j$ types of colloids, and the values 1 and 2 indicate small and large colloids, respectively. Figure~\ref{grmax} plots their peak values $g_{max,11}$, $g_{max,12}$, and $g_{max,22}$ as a function of $\phi$ at various $\rho_r$, indicating that all of them increase monotonously in the range of $\phi$ considered in this study. This behavior, reminiscent of hard spheres, shows that standard feature of charged colloids may be mapped onto an equivalent hard sphere system with a much larger effective diameter~\cite{Rosenberg86PRA}.

{\bf Appendix E: Individual fits to data in Figure~\ref{RFOT_total} (C)}

\begin {figure}
\centering
\includegraphics  [width=3.5in] {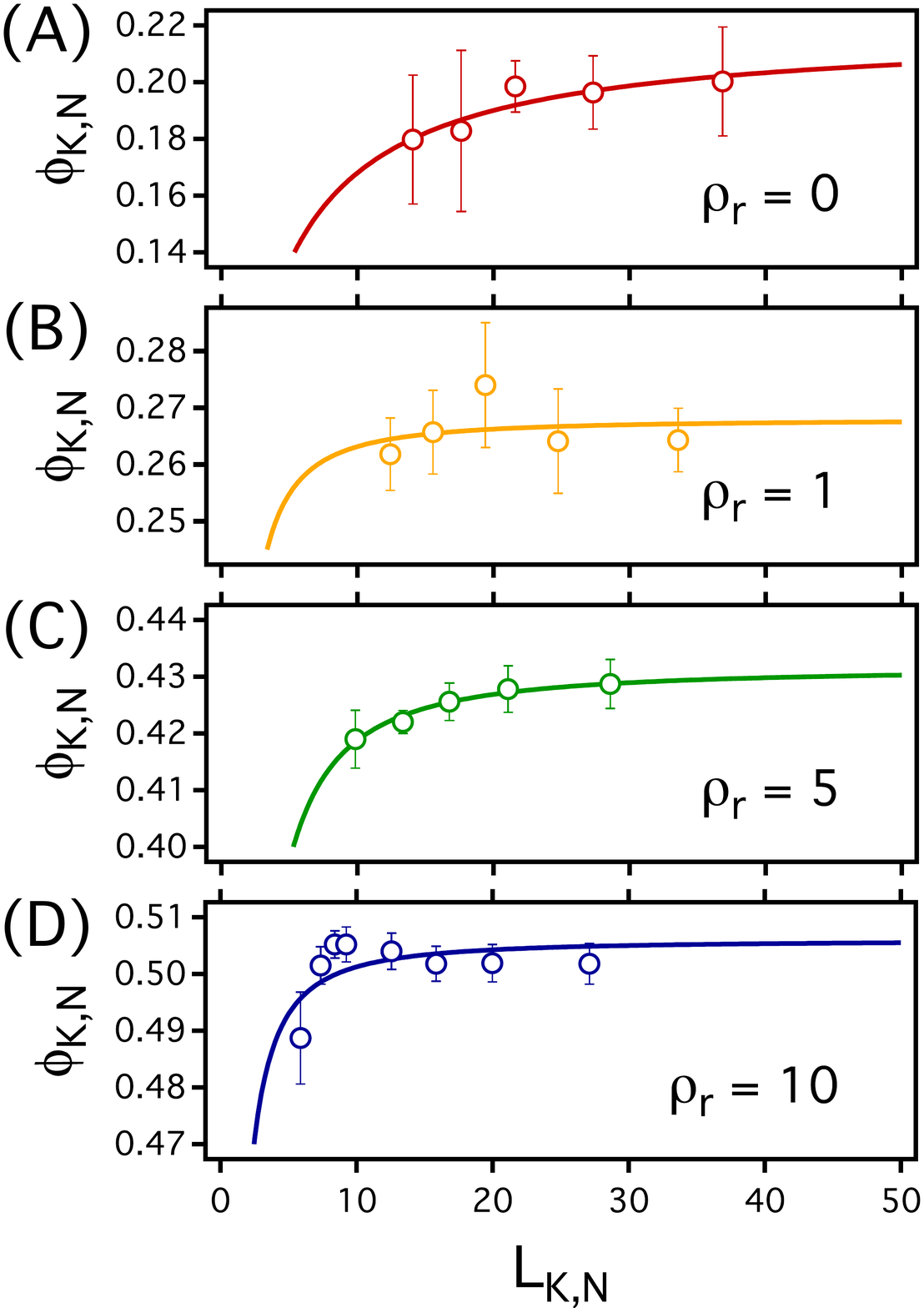}
\caption{Individual fits of Figure~\ref{RFOT_total} (C). $\phi_{K,N}$ is $\phi_K$ at $N$, which is obtained by the VFT fit as in Figure~\ref{Dynamic} (C). In the each panel, $\phi_{K,N}$ of $L_{K,N}$ (the open circles) is fitted to Eq~(\ref{FSS1}) (the solid lines), by which $\phi_{K,\infty}$ and $L_0$ are obtained as fitting parameters.}
\label{RFOT_supple}
\end{figure}

In Figure~\ref{RFOT_supple}, we show how $\phi_{K,N}$ changes with $L_{K,N}$ as a function of $N$. The value of $L_{K,N}$ is given by $(N.\phi_{K,N})^{1/3}$. The $\rho_r$ values are shown in Figure~\ref{RFOT_supple}.

%


\end{document}